\documentclass[aip, amsmath,amssymb,reprint]{revtex4-1}
\usepackage{dcolumn}
\usepackage{bm}
\usepackage[utf8]{inputenc}
\usepackage[T1]{fontenc}
\usepackage{mathptmx}

\usepackage{lipsum}
\usepackage{indentfirst}
\usepackage{fancyhdr}
\usepackage{amssymb}
\usepackage{amsmath}
\usepackage{url}
\usepackage{setspace}
\usepackage{graphicx,color}
\usepackage{fancyvrb }
\usepackage{float}
\usepackage{enumerate}
\usepackage{textcomp}
\usepackage{nomencl}
\usepackage{comment}
\usepackage[acronym]{glossaries}
\usepackage{tabto}
\usepackage{appendix}
\usepackage{upgreek}
\usepackage{mathtools}
\usepackage{gensymb}
\usepackage{hyperref}
\usepackage{dblfloatfix}
\numberwithin{equation}{section}
\numberwithin{figure}{section}
\DeclareMathOperator\erf{erf}

\begin{document}



\title[Optimisation of Thin Plastic Foil Targets for Production of Laser-Generated Protons in the GeV Range]{Optimisation of Thin Plastic Foil Targets for Production of Laser-Generated Protons in the GeV Range}



\author{P. Hadjisolomou}
\email[Electronic mail: ]{Prokopis.Hadjisolomou@eli-beams.eu}
\affiliation{Institute of Physics of the ASCR, ELI-Beamlines, Na Slovance 2, 18221 Prague, Czech Republic}

\author{I. P. Tsygvintsev}
\email[Electronic mail: ]{IliaTsygvintsev@gmail.com}
\altaffiliation[Also at ]{RnD-ISAN/EUV Labs, Promyshlennaya Str., 1A, Moscow-Troitsk, 142191, Russia}
\affiliation{Keldysh Institute of Applied Mathematics RAS, Miusskaya sq., 4, Moscow, 125047, Russia}

\author{P. Sasorov}
\email[Electronic mail: ]{Pavel.Sasorov@eli-beams.eu}
\altaffiliation[Also at ]{Keldysh Institute of Applied Mathematics RAS, Miusskaya sq., 4, Moscow, 125047, Russia}
\affiliation{Institute of Physics of the ASCR, ELI-Beamlines, Na Slovance 2, 18221 Prague, Czech Republic}

\author{V. Gasilov}
\email[Electronic mail: ]{vgasilov@keldysh.ru}
\affiliation{Keldysh Institute of Applied Mathematics RAS, Miusskaya sq., 4, Moscow, 125047, Russia}

\author{G. Korn}
\email[Electronic mail: ]{Georg.Korn@eli-beams.eu}
\affiliation{Institute of Physics of the ASCR, ELI-Beamlines, Na Slovance 2, 18221 Prague, Czech Republic}

\author{S. V. Bulanov}
\email[Electronic mail: ]{Sergei.Bulanov@eli-beams.eu}
\altaffiliation[Also at ]{Kansai Photon Science Institute, 8-1-7 Umemidai, Kizugawa, Kyoto 619-0215, Japan}
\affiliation{Institute of Physics of the ASCR, ELI-Beamlines, Na Slovance 2, 18221 Prague, Czech Republic}


\date{\today}

\begin{abstract}
In order to realistically simulate the interaction of a femtosecond laser pulse with a nanometre-thick target it is necessary to consider a target preplasma formation due to the nanosecond long amplified-spontaneous-emission pedestal and/or prepulse. The relatively long interaction time dictated that hydrodynamic simulations should be employed to predict the target particles' number density distributions prior the arrival of the main laser pulse. By using the output of the hydrodynamic simulations as input into particle-in-cell simulations, a detailed understanding of the complete laser-foil interaction is achieved. Once the laser pulse interacts with the preplasma it deposits a fraction of its energy on the target, before it is either reflected from the critical density surface or transmitted through an underdense plasma channel. A fraction of hot electrons is ejected from the target leaving the foil in a net positive potential, which in turn results in proton and heavy ion ejection. In this work protons reaching ${\sim} \kern0.1em 25 \kern0.2em \mathrm{MeV}$ are predicted for a laser of ${\sim} \kern0.1em 40  \kern0.2em \mathrm{TW}$ peak power and ${\sim} \kern0.1em 600  \kern0.2em \mathrm{MeV}$ are expected from a ${\sim} \kern0.1em 4  \kern0.2em \mathrm{PW}$ laser system.
\end{abstract}


\maketitle 


\section{Introduction}

\par The recent development of laser technology leads to ${\sim} \kern0.1em 10  \kern0.2em \mathrm{PW}$ laser availability. These lasers are expected to be one of the major tools regarding the study of laser-matter interaction. Furthermore, proton acceleration schemes that promise ${\sim} \kern0.1em \mathrm{GeV}$ energies are theoretically formulated \cite{Esirkepov2004} and be tested for relatively low proton energies. On the other hand, near-$\mathrm{PW}$ facilities have already produced ${\sim} \kern0.1em100  \kern0.2em \mathrm{MeV}$ protons \cite{Higginson2018}.

\par Proton acceleration has attracted significant attention over the past decades, due to the numerous promising and novel applications it can be applied to. Some of these applications include nuclear isotope production \cite{Santala2001}, material science \cite{Booty1996}, proton radiography \cite{Borghesi2002, Borghesi2003} and fast ignition \cite{Roth2001, Atzeni2002}. Among all the proposed applications, cancer therapy \cite{Bulanov2002, Bulanov2004} has a significant importance but also requires a proton energy of ${\sim} \kern0.1em 250 \kern0.2em \mathrm{MeV}$, which is significantly higher than the current record.

\par The basic idea behind laser-generated proton acceleration is that when a ${\sim} \kern0.1em \mathrm{PW}$ laser pulse is focused onto a spot of a few micrometres radius, then the laser intensity is so high that electrons are ejected from the target, leaving the target in a net positive potential which in turn results on proton ejection. As described in Refs. \onlinecite{Bulanov2004, Bulanov2014}, the maximum proton energy is given by $\varepsilon_p \approx 173 \sqrt{P[\mathrm{PW}]} \kern0.2em \mathrm{MeV}$, where $P[\mathrm{PW}]$ is the laser power given in $\mathrm{PW}$. By considering three lasers of $0.1 \kern0.2em \mathrm{PW}$, $1 \kern0.2em \mathrm{PW}$ and $10 \kern0.2em \mathrm{PW}$, the expression for the maximum proton energy yields $55 \kern0.2em \mathrm{MeV}$, $173 \kern0.2em \mathrm{MeV}$ and $547 \kern0.2em \mathrm{MeV}$ protons respectively.

\par This work considers pulses originating from a typical Ti:Sapphire laser \cite{Mourou2006}. Such a pulse can be realised as the sum of several pulse components, such as the main femtosecond pulse (${\sim} \kern0.1em 30 \kern0.2em \mathrm{fs}$), several prepulses (${\sim} \kern0.1em 30 \kern0.2em \mathrm{fs}$), an amplified spontaneous emission (ASE) pedestal (${\sim} \kern0.1em 1 \kern0.2em \mathrm{ns}$) and a post-pulse (${\sim} \kern0.1em 0.1 \kern0.2em \mathrm{ns}$). The contrast of the laser pulse is defined as the ratio of the main pulse amplitude to the ASE pedestal amplitude and it usually has a value near $10^{10}$. The contrast value mainly depends on whether or not a plasma mirror is used in the laser system, allowing a flexibility on the contrast value choice \cite{Mourou2006, Levy2007}.

\par As it is realised theoretically in Refs. \onlinecite{Matsukado2003, Yogo2008, Esirkepov2014} and experimentally in Refs. \onlinecite{Matsukado2003, Yogo2008, Ogura2012}, the ASE pedestal heats an initially steep density gradient flat-foil target, resulting in a modification of the target's number density distribution. Although the extent of this effect depends on the ASE pedestal intensity, focal spot, pulse duration and foil thickness/density, in all cases the modified target is curved in the vicinity of the focal spot region, characterised by a finite (smooth) density gradient. Therefore, when the main pulse arrives on the target it faces completely different initial conditions than what were initially assumed (by the absence of the ASE pedestal), modifying the laser-foil interaction. As an extension, the resulting proton/ion spectra are significantly different than those resulting from steep density gradient flat targets, where as indicated in the literature \cite{Matsukado2003, Yogo2008, Esirkepov2014,  Fuchs2007,  McKenna2008}, the existence of a large preplasma gradient in general benefits the proton acceleration. As for the effect of the ${\sim} \kern0.1em 30 \kern0.2em \mathrm{fs}$ prepulses can be ignored due to their extremely small duration (compared to the ASE pedestal), while the post-pulse effect can also be ignored since by its' arrival time the main pulse has already interacted with the target.

\par The main aim of this paper is to study the laser-foil interaction and the resulting proton/ion acceleration under the conditions of a modified target geometry caused by the ASE pedestal. Due to the long simulation time needed ($1 \kern0.2em \mathrm{ns}$) to model the finite contrast effects, this work is realised as the combination of two methods. Initially, hydrodynamic modelling is employed which studies the effect of the ASE pedestal and gives the modified target distribution prior the arrival of the main pulse. By using that distribution as an initial condition, particle-in-cell (PIC) simulations are employed, which simulate the interaction of the main pulse with the ASE pedestal modified target. A multi-parametric study of the interaction is performed by combining simulation cases with target thicknesses of $0.1 \kern0.2em \mathrm{\upmu m}$, $0.3 \kern0.2em \mathrm{\upmu m}$ and $0.9 \kern0.2em \mathrm{\upmu m}$, ASE pedestal intensities of $ 0 \kern0.2em \mathrm{W cm^{-2}}$ (undisturbed targets), ${ 1.6 \kern0.1em {\times} \kern0.1em 10^{9} \kern0.2em \mathrm{W cm^{-2}} }$ and ${ 1.6 \kern0.1em {\times} \kern0.1em 10^{10} \kern0.2em \mathrm{W cm^{-2}} }$, and main pulse intensities of ${ 1.85 \kern0.1em {\times} \kern0.1em 10^{20} \kern0.2em \mathrm{W cm^{-2}} }$, ${ 1.85 \kern0.1em {\times} \kern0.1em  10^{21} \kern0.2em \mathrm{W cm^{-2}} }$ and ${ 1.85 \kern0.1em {\times} \kern0.1em  10^{22} \kern0.2em \mathrm{W cm^{-2}} }$. Although our results in general agree with that the existence of a preplasma benefits the proton acceleration, a few cases are identified where the opposite condition occurs.

\par This paper is organised in four sections. In Sec. \ref{Background} a detailed description of the main pulse is given as used in the PIC simulations, along with a quantification of the relevant physical parameters. Additionally, a brief description of the hydrodynamic modelling is presented. Based on the results of the hydrodynamic model, a characterisation of the electron distributions used in the simulations is also made. Sec. \ref{Setup} presents a detailed explanation of the parameters used the PIC simulations. The main results of the paper are presented in Sec. \ref{Results}, where the outcome of the parametric study is presented; Sec. \ref{Results} further splits in four subsections. Subsec. \ref{absorption} indicates the energy fraction transferred from the laser pulse to the particles. Subsec. \ref{field} mentions the behaviour of the remaining fraction of the pulse after the interaction with the target. Subsec. \ref{Electron Spectra} separates the resulting hot electrons in several distinct populations. Subsec. \ref{Proton Spectra} presents the multi-parametric study from the perspective of proton/ion acceleration and a direct comparison of all cases is made, extracting crucial results which allow the optimisation of the proton energy in the three laser intensity levels examined. The paper ends with a conclusive section presented in Sec. \ref{Conclusions}.


\section{Theoretical Background} \label{Background}


\subsection{Laser Pulse Characterisation} \label{characterisation}

\begin{figure}[!ht]
  \includegraphics[width=0.9\linewidth]{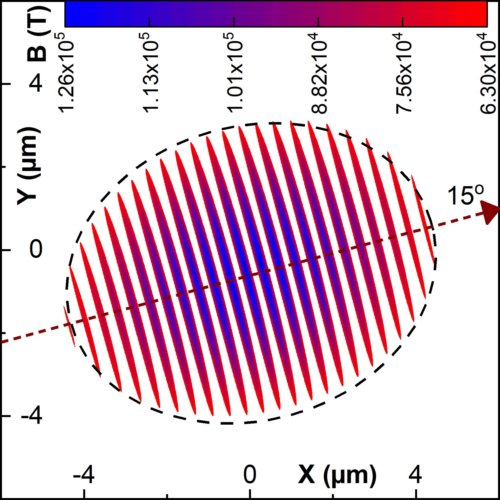}
  \caption{The magnetic field of a linearly polarised laser pulse propagating in vacuum, as extracted from the EPOCH \cite{Arber2015} code. The pulse has a peak intensity of $1.85 \kern0.1em {\times} \kern0.1em 10^{20} \kern0.2em \mathrm{W cm^{-2}}$, focal spot of $5 \kern0.2em \mathrm{\upmu m}$ pulse duration of $30 \kern0.2em \mathrm{fs}$ and wavelength of $810 \kern0.2em \mathrm{nm}$. The pulse propagates on the ${\bold{\hat{z}}}$ plane and forms an angle of $15 \degree$ (brown dashed arrow) with the ${\bold{\hat{y}}}$ plane. The dashed black elliptic line represents the expected extent of the pulse at the FWHM, as described within this subsection. For a direct comparison, the magnetic field scale of the pulse starts from half its' peak value.}
  \label{fig:pulse_sample}
\end{figure}

\par One of the aims of this work is the optimisation of a planned experiment in the ELI-Beamlines, Czech Republic, where nanometre thick Mylar \cite{DeMeuse2011} foils can be used at an intensity equal the lowest out of three presently examined. Therefore, the precise definition of the pulse parameters is necessary in order to predict the expected proton spectra. The laser energy, $E_0$, contained within the first minimum of diffraction after focusing of the laser pulse is expected to be $1.31 \kern0.2em \mathrm{J}$ with a focal spot diameter full width at half maximum (FWHM), $d_{FWHM}$, of $5 \kern0.2em \mathrm{\upmu m}$ and a pulse duration FWHM, $t_{FWHM}$, of $30 \kern0.2em \mathrm{fs}$. The pulse is P-polarised, its' angle of incidence on the target (with respect to target normal) equals $15 \degree$ and its' wavelength, $w$, is of $810 \kern0.2em \mathrm{nm}$.

\par The pulse temporal and spatial profiles can be assumed as Gaussians of the form $(\sigma \sqrt{2 \kern0.1em \pi})^{-1} \exp[-(x/(\sqrt{2} \kern0.1em \sigma))^2]$, where $\sigma$ represents the standard deviation, associated with the FWHM as $2 \sqrt{2 \ln(2)} \kern0.1em \sigma$. For a Gaussian beam, the fraction of its' area contained within $\pm m \kern0.1em \sigma$ from its' centroid (where m is a real number) is given by $\erf(m/\sqrt{2})$, where $\erf$ is the error function \cite{Houston2012}. The intensity, $I$, of a Gaussian beam peaks at the centroid region. The Gaussian beam intensity approaches the intensity of a flat-top beam at a distance $m \kern0.1em \sigma$ from the centroid, where now m is a small integer. Therefore, one can write:
\begin{equation}
I = \frac{E(m)} { {2 \kern0.1em t_{\sigma}(m)} \left [\pi \kern0.1em r^2_{\sigma}(m) \right]} \kern0.2em ,
\label{I1}
\end{equation}
where the subscript $\sigma$ denotes that the duration, $t_{\sigma}$, and radius, $r_{\sigma}$ are associated with the standard deviation. The factor of 2 on the denominator is because the pulse duration under the $\sigma$-representation needs to correspond to both positive and negative values of $m$ relatively to the centroid. The intensity in Eq. \ref{I1} is defined by three Gaussians (one temporal and two spatial). Therefore, by considering the amount of energy associated with $m$, one can write Eq. \ref{I1} as:
\begin{equation}
I = \frac{E_0} { {2 \kern0.1em m \kern0.1em t_{\sigma}} \left [ \pi (m \kern0.1em r^2_{\sigma}) \right]} \left[ \erf \left( \frac{m}{\sqrt{2}} \right) \right]^3 \kern0.2em .
\label{I2}
\end{equation}
\vspace{1mm}

\par The peak intensity, $I_{max}$, is obtained to the limit of Eq. \ref{I2} as $m \rightarrow 0$, giving:
\begin{equation}
I_{max} = \frac{E_0} { {2 \kern0.1em t_{\sigma}} (\pi \kern0.1em r^2_{\sigma})} \left( \frac{2}{\pi} \right)^{3/2} \kern0.2em ,
\label{I3}
\end{equation}
or, if expressed in the most commonly used terms of FWHM it gets the form:
\begin{equation}
I_{max} = \frac{E_0} { {t_{FWHM}} \left [ \pi (d_{FWHM}/2)^2 \right ] } \left( \frac{4 \ln(2)}{\pi} \right)^{3/2} \kern0.2em .
\label{I4}
\end{equation}
For the above specified parameters, Eq. \ref{I4} gives a peak intensity of $1.85 \kern0.1em {\times} \kern0.1em 10^{20} \kern0.2em \mathrm{W cm^{-2}}$ \kern0.2em .


\begin{figure}
  \includegraphics[width=\linewidth]{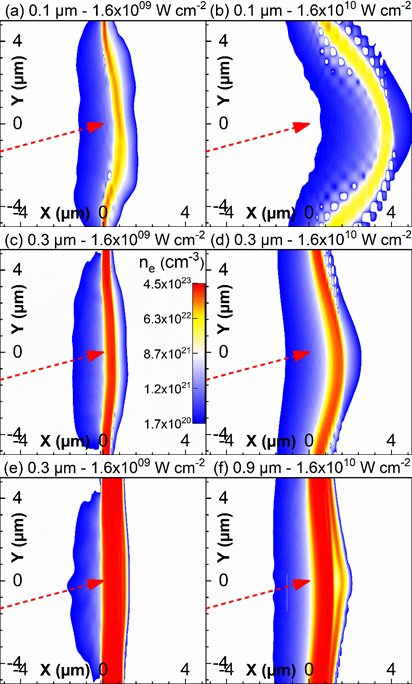}
  \caption{Electron number density on the ${\bold{\hat{z}}}$ plane (at $z=0$), due to the act of a $1 \kern0.2em \mathrm{ns}$ ASE pedestal on a flat-foil with an intensity of ${ 1.6 \kern0.1em {\times} \kern0.1em 10^{9} \kern0.2em \mathrm{W cm^{-2}} }$ and ${ 1.6 \kern0.1em {\times} \kern0.1em 10^{10} \kern0.2em \mathrm{W cm^{-2}} }$ for the left and right column respectively. The foils hve an initially steep density gradient and a uniform density, where the initial foil thickness is $0.1 \kern0.2em \mathrm{\upmu m}$, $0.3 \kern0.2em \mathrm{\upmu m}$ and $0.9 \kern0.2em \mathrm{\upmu m}$ for the first, second and third row respectively. The laser propagation axis ($15 \degree$) is represented by the red dashed arrow.}
  \label{fig:preplasma}
\end{figure}


\par The corresponding peak electric field magnitude, ${{E}_{max}}$, can be calculated by the expression:
\begin{equation}
{{E}_{max}} = \sqrt{\frac{2 \kern0.1em I_{max}}{\varepsilon_0 \kern0.1em c}} \kern0.2em ,
\label{E1}
\end{equation}
where $\varepsilon_0$ is the electric permittivity in free space and $c$ is the vacuum speed of light. Eq. \ref{E1} gives a peak electric field of $3.73 \kern0.1em {\times} \kern0.1em 10^{13} \kern0.2em \mathrm{V m^{-1}}$ for the above specified intensity, while it gives an order of magnitude higher field for an intensity of $1.85 \kern0.1em {\times} \kern0.1em 10^{22} \kern0.2em \mathrm{W cm^{-2}}$, in agreement with Fig. \ref{fig:pulse_sample}. The FWHM of the electric field is related to the laser focal spot diameter as $\sqrt{2} \kern0.1em d_{FWHM}$, giving a ${\sim7 \kern0.2em \mathrm{\upmu m}}$ extent for a laser focal spot of $5 \kern0.2em \mathrm{\upmu m}$. Furthermore, a $30 \kern0.2em \mathrm{fs}$ pulse duration corresponds to a spatial extent of $9 \kern0.2em \mathrm{\upmu m}$, which corresponds to the dashed black ellipsis drawn in Fig. \ref{fig:pulse_sample}.


\subsection{Hydrodynamic Preplasma Formation} \label{hydro}

\par Simulations estimating the preplasma formation are carried in a full three-dimensional geometry using the hydrodynamical approach. The simulation tool used is the 3DLINE \cite{Tsygvintsev2017} code, designed for radiative laser plasma simulations. The code is used alongside with a one-fluid one-temperature quasi-neutral model of plasma with a constant chemical composition. The set of differential equations governing motion and heat transfer can be written in the form:
\setlength{\jot}{12pt}
\begin{equation}
\begin{aligned}
  \frac{d \rho}{dt} &=-\rho \kern0.1em \nabla \cdot \bold{u} \kern0.2em , \\
  \rho \kern0.1em \frac{d\bold{u}}{dt}&= -\nabla P \kern0.2em , \\
  \frac{d\varepsilon}{dt}&=-P \kern0.1em \frac{d(1/\rho)}{dt} - \frac{1}{\rho} \nabla \cdot \bold{W} + G^{rad} + G^{las} \kern0.2em ,
\end{aligned}
  \label{3DLINE:eq1}
\end{equation}
where ${d}/{dt} = \partial_t + ({\bf{u}} \kern0.1em \nabla$) is the full (substantial) time derivative, $\rho$ is the matter density, $\bold{u}$ is the velocity of the flow, $P$ is the pressure, $\varepsilon$ is the specific internal energy, $\bold{W}=-\kappa \kern0.1em \nabla T$ is the thermal flux, $\kappa$ is the thermal conductivity coefficient, $T$ is the temperature and $G^{rad}$, $G^{las}$ are specific sources (sinks) of energy due to radiation transfer and laser power deposition. In order to determine these sources, additional equations need to be used.

\par The thermal radiation transfer is calculated using the one-group diffusion approximation:
\begin{equation}
\begin{aligned}
\rho \kern0.1em G^{rad}&= -Q + c \kern0.1em \varkappa^P \kern0.1em U \kern0.2em , \\
\nabla\cdot {\bold{W}}^{rad}&= Q-c \kern0.1em \varkappa^P \kern0.1em U \kern0.2em , \\
{\bold{W}}^{rad}&=-\frac{c}{3 \kern0.1em \varkappa^R} \kern0.1em \nabla U \kern0.2em .
\end{aligned}
  \label{3DLINE:eq2}
\end{equation}
Here, $Q$ is the total volumetric energy loss due to radiation, $c$ is the speed of light in vacuum, $\varkappa^P$ is the Planck's opacity, $\varkappa^R$ is the Rosseland's opacity, $U$ is the radiation energy density and ${\bold{W}}^{rad}$ is the radiation energy flux. The optical properties as a function of the temperature and density are calculated with the THERMOS code \cite{Nikiforov2005}. While radiation anisotropy in the preplasma is beyond the scope of the diffusion approximation, it is fully applicable at the ablation layer, where radiation flux can influence the target dynamics. However, since there is no heavy element at the target compound (hydrogen, carbon and oxygen), this factor is relatively low; in all simulated cases the integral laser energy conversion to radiation is ${\sim} \kern0.1em 1 \kern0.2em \%$.

\par For simulation of the laser energy transfer and deposition a ``hybrid'' model \cite{Tsygvintsev2016,Basko2017} is used. This model combines a three-dimensional ray tracing in the geometrical optics approximation \cite{Kaiser2000} with a one-dimensional solution of the Helmholtz equation \cite{Born1980}. The integral absorption of laser energy is ${\sim} \kern0.1em 70 \kern0.2em \%$ for an ASE prepule intensity of ${ 1.6 \kern0.1em {\times} \kern0.1em 10^{9} \kern0.2em \mathrm{W cm^{-2}} }$ and ${\sim} \kern0.1em 90 \kern0.2em \%$ for $1.6 \kern0.1em {\times} \kern0.1em 10^{10} \kern0.2em \mathrm{W cm^{-2}}$.

\par The equation of state (which couples the pressure and internal energy with the temperature and density) is calculated for a chemical mix of constant compound using the FEOS \cite{Faik2018} code, based on Thomas-Fermi's approximation with half-empiric corrections \cite{Kemp1998}. At the phase transition region the Maxwell's construction is applied, leading to a phase equilibrium. Although this approach can not describe an overheated liquid phase state, which occurs at ASE pedestal timescales of ${\sim} \kern0.1em 1 \kern0.2em \mathrm{ns}$, it can still give a quantitatively correct estimation for the target mass dynamics at the target rear side \cite{Basko2018}.

\par A spatial discretisation of equations \eqref{3DLINE:eq1} and \eqref{3DLINE:eq2} is done on the staggered grid; thermodynamical properties ($\rho$, $T$, $P$, $\varepsilon$, etc.), velocities $\bold{u}$ and fluxes (${\bold{W}}$ and ${\bold{W}}^{rad}$) are approximated to cells, points and facets respectively. This approach allows the easy construction of divergence form of equations, providing conservation laws for mass, energy and momentum. It also provides second order accuracy of spatial approximation for the work of the pressure term $P \kern0.1em d(1/\rho)$ on the regular grid \cite{Samarskii1975}.

\par For the calculation of convective fluxes of mass ${(\bold{u} \kern0.1em {\cdot} \kern0.1em \nabla) \kern0.1em \rho}$ and internal energy $(\bold{u} \kern0.1em {\cdot} \kern0.1em \nabla) \kern0.1em \rho \kern0.1em\varepsilon$ a second-order piecewise parabolic method \cite{Colella1984} is used. The convective fluxes of momentum and kinetic energy are matched with the method described in Ref. \onlinecite{Gasilov2017} using a half-explicit approximation. This method modification breaks the complete conservativity of the difference scheme with an integral energy error of ${{\sim} \kern0.1em \Delta t^3}$. However, the obtained equations are linear,  in contrary with the original approach. The main downside of the described hydrodynamical scheme is the required timestep restriction; to completely suppress unphysical oscillations it is necessary to set Courant's number as low as $0.1$. While for the estimation of the thermal conductivity a completely implicit scheme is used, source terms are calculated with an explicit approach.

\par The calculations are performed on a rectilinear (${\bold{\hat{x}}}, {\bold{\hat{y}}}, {\bold{\hat{z}}}$) grid, with $520 \kern0.1em {\times} \kern0.1em 40 \kern0.1em {\times} \kern0.1em 20$ cells corresponding to a volume of $20 \kern0.1em {\times} \kern0.1em 20 {\times} \kern0.1em 10 \kern0.2em \mathrm{\upmu m}$. The laser pulse axis is lying on the ${\bold{\hat{z}}}$ plane (at $z = 0$), forming an angle of $15 \degree$ with the ${\bold{\hat{x}}}$ axis, as shown in Fig. \ref{fig:pulse_sample}. The lower cell number along the ${\bold{\hat{z}}}$ axis is due to the reflection symmetry at $z=0$. The space discretization along ${\bold{\hat{y}}}$ and ${\bold{\hat{z}}}$ is uniform, with a spatial step of $0.5 \kern0.2em \mathrm{\upmu m}$. The discretization along ${\bold{\hat{x}}}$ is significantly refined in order to provide a good resolution of the target dynamics inside the flat-foil; the step is $0.01 \kern0.2em \mathrm{\upmu m}$ at a $1 \kern0.2em \mathrm{\upmu m}$ region covering the initial position of the target, and then the step is increasing in a geometrical progression up to a value of $0.1 \kern0.2em  \mathrm{\upmu m}$ at the simulation borders along ${\bold{\hat{x}}}$.


\subsection{Altered Target Density Distribution} \label{altered}

\begin{figure}[!ht]
  \includegraphics[width=\linewidth]{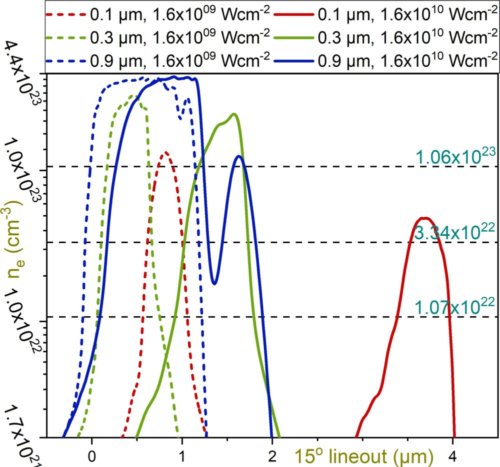}
  \caption{Line-out of the preplasma distribution (shown in Fig. \ref{fig:preplasma}) along the laser axis ($15 \degree$ to target normal). The solid and dashed lines correspond to an ASE pedestal intensity of $1.6 \kern0.1em {\times} \kern0.1em 10^9 \kern0.2em \mathrm{W cm^{-2}}$ and ${ 1.6 \kern0.1em {\times} \kern0.1em 10^{10} \kern0.2em \mathrm{W cm^{-2}} }$ respectively. The red, green and blue colour corresponds to a target thickness of $0.1 \kern0.2em \mathrm{\upmu m}$, $0.3 \kern0.2em \mathrm{\upmu m}$ and $0.9 \kern0.2em \mathrm{\upmu m}$ respectively. The baseline of the graph equals the classical critical density for the laser pulse parameters presented in this paper. The dashed lines at ${1.07 \kern0.1em {\times} \kern0.1em 10^{22} \kern0.2em \mathrm{cm^{-3}} }$, $3.34 \kern0.1em {\times} \kern0.1em 10^{22} \kern0.2em \mathrm{cm^{-3}}$ and $1.06 \kern0.1em {\times} \kern0.1em 10^{23} \kern0.2em \mathrm{cm^{-3}}$ correspond to the relativistically corrected critical densities for main pulse peal intensities of ${ 1.85 \kern0.1em {\times} \kern0.1em 10^{20} \kern0.2em \mathrm{W cm^{-2}} }$, $1.85 \kern0.1em {\times} \kern0.1em 10^{21} \kern0.2em \mathrm{W cm^{-2}}$ and $1.85 \kern0.1em {\times} \kern0.1em 10^{22} \kern0.2em \mathrm{W cm^{-2}}$ respectively.}
\label{fig:preplasma_lineout_15deg}
\end{figure}

\par All the hydrodynamic simulations presented in this work have a fixed ASE pedestal duration of $1 \kern0.2em \mathrm{ns}$ and a focal spot of $5 \kern0.2em \mathrm{\upmu m}$. Although the ASE pedestal duration strongly alters the preplasma gradient (the focal spot mainly affects the extent of preplasma on the target surface), it is considered worth altering only the ASE pedestal intensity rather than trying to investigate every possible combination in order to be able to apply a direct comparison to the simulations' outcome. Two sets of simulations are carried out, based on ASE pedestal intensities of $1.6 \kern0.1em {\times} \kern0.1em 10^9 \kern0.2em \mathrm{W cm^{-2}}$ and $1.6 \kern0.1em {\times} \kern0.1em 10^{10} \kern0.2em \mathrm{W cm^{-2}}$. If a main pulse of $1.85 \kern0.1em {\times} \kern0.1em 10^{20} \kern0.2em \mathrm{W cm^{-2}}$ is considered, the corresponding contrast ratios correspond to a value of ${ {\sim} \kern0.1em 10^{11} }$ and ${\sim} \kern0.1em 10^{10}$ respectively. Although these contrast values sound extremely optimistic from an experimental point of view, their value can be compensated from a shorter ASE pedestal duration which if desired can be several times less than the $1 \kern0.2em \mathrm{ns}$ assumed \cite{Sung17}. Emphasis is given to obtaining the appropriate preplasma distributions that result on different proton acceleration characteristics, as shown in Fig. \ref{fig:preplasma}. Note that the filament-like structures appearing  in Fig. \ref{fig:preplasma}(b) are unphysical artefacts due to interpolation of the hydrodynamic simulation output into a $2000 \kern0.1em {\times} \kern0.1em 2000$ array.

\par For all simulations a Mylar foil is assumed, having an electron number density of $4.38  \kern0.1em {\times} \kern0.1em 10^{23}  \kern0.2em \mathrm{cm^{-3}}$. Studying of the proton acceleration on the properties of the target material is beyond the scope of the present work. The target thickness is another parameter (along with the ASE pedestal intensity) affecting the preplasma formation, where as an extension, this work examines how the proton acceleration process is affected by the target thickness for main pulse intensities in the range of $1.85 \kern0.1em {\times} \kern0.1em 10^{20} \kern0.2em \mathrm{W cm^{-2}}$ - $1.85 \kern0.1em {\times} \kern0.1em 10^{22} \kern0.2em \mathrm{W cm^{-2}}$.


\par The equal density preplasma contours in Fig. \ref{fig:preplasma} are summarised in Fig. \ref{fig:preplasma_lineout_15deg}, where a line-out along the laser propagation axis ($15 \degree$) is shown for all preplasma distributions examined in this work. The dashed lines correspond to a ${ 1.6 \kern0.1em {\times} \kern0.1em 10^9 \kern0.2em \mathrm{W cm^{-2}} }$ ASE pedestal intensity and the solid lines to $1.6 \kern0.1em {\times} \kern0.1em 10^{10} \kern0.2em \mathrm{W cm^{-2}}$. The difference in colour corresponds to difference in foil thickness, where red is for $0.1 \kern0.2em \mathrm{\upmu m}$, green is for $0.3 \kern0.2em \mathrm{\upmu m}$ and blue is for $0.9 \kern0.2em \mathrm{\upmu m}$ thick foils. The base-line of the graph corresponds to the classical critical density and the three dashed horizontal lines correspond to relativistically corrected critical densities for the three main pulse intensities used in this work (from lower to higher).

\par From Fig. \ref{fig:preplasma_lineout_15deg} it is seen that a different extent of preplasma formation exists, which depends on both the ASE pedestal intensity and the foil thickness. In general, for a higher contrast ratio the preplasma formation is less and the initially steep target density gradient tends to keep its' original slope. For the ASE pedestal intensity of ${ 1.6 \kern0.1em {\times} \kern0.1em 10^9 \kern0.2em \mathrm{W cm^{-2}} }$ the target density envelope has approximately a Gaussian/Super-Gaussian form, while for an order of magnitude higher intensity the density distribution transforms to a Skewed Gaussian (with the addition of a lower magnitude Gaussian for the $0.9 \kern0.2em \mathrm{\upmu m}$ thick foil). Furthermore, the target density distributions for a higher contrast ASE pedestal are closer to the original foil location, as the dashed line distributions in Fig. \ref{fig:preplasma_lineout_15deg} are located relatively near to the axis origin, in contrast to the solid line distributions. This effect can be better seen in Fig. \ref{fig:preplasma}, where an extensive foil curvature is observed for the lower contrast ratio cases. However, the foil is not deformed out in the laser pulse interaction region, resulting in much different proton trajectories, as seen in section \ref{Results}.

\par The second factor affecting the preplasma distribution is the foil thickness. Although the target density distribution is mostly affected by the contrast ratio, the extent of this alteration is strongly affected by the target thickness. In general, the change in the FWHM of the distribution is significantly higher for thinner targets, where for the $0.1 \kern0.2em \mathrm{\upmu m}$ thick foil the ratio of the final to initial FWHM is ${\sim} \kern0.1em 2.8$ for the $1.6 \kern0.1em {\times} \kern0.1em 10^9 \kern0.2em \mathrm{W cm^{-2}}$ ASE pedestal intensity; the same quantity has a value of ${\sim} \kern0.1em 1.4$ and ${\sim} \kern0.1em 1.2$ if the foil thickness is increased to $0.3 \kern0.2em \mathrm{\upmu m}$ and $0.9 \kern0.2em \mathrm{\upmu m}$ respectively. Additionally, the peak of the density distribution drops by reducing the foil thickness, resulting in a drastically different proton acceleration behaviour, as described in section \ref{Results}.


\section{PIC Simulations Set-Up} \label{Setup}

\par In this work the PIC code EPOCH \cite{Arber2015} is used in the two-dimensional (2D) version. The code is compiled with the Higuera-Cary \cite{Higuera2017, Ripperda2018} flag enabled (by default the Boris solver \cite{Ripperda2018, Boris1970} is used), which gives more accurately the ${ {\bf{E}} \kern0.1em {\times} \kern0.1em {\bf{B}} }$ velocity, while at the same time is volume-preserving. The use of the Higuera-Cary solver becomes more important at higher velocities, where as an example $150 \kern0.2em \mathrm{MeV}$ protons have a velocity above $0.5 \kern0.2em c $.

\par The simulations run on the ECLIPSE cluster on $36$ nodes (with $16$ processors in each node) resulting in a 2D processor rearrangement of $24 \kern0.1em {\times} \kern0.1em 24$. The simulation has a simulation box of $122.88 \kern0.2em \mathrm{\upmu m} \kern0.1em {\times} \kern0.1em 122.88 \kern0.2em \mathrm{\upmu m}$ with $3 \kern0.1em {\times} \kern0.1em 10^{12}$ cells in each direction, resulting in a cell size of $10 \kern0.2em \mathrm{nm}$; this value is approximately half of the relativistically corrected skin depth for Mylar assuming the pulse intensity of $1.85 \kern0.1em {\times} \kern0.1em 10^{20} \kern0.2em \mathrm{W cm^{-2}}$ ($20.7 \kern0.2em \mathrm{nm}$). A total number of $9 \kern0.1em {\times} \kern0.1em 10^{25}$ macroparticles (two times the total number of cells) per particle specie is used; by considering the number of empty cells in the simulation it is estimated that each cell initially contains ${\sim} \kern0.1em 2^7$ macroparticles, with the exact number depending on the foil thickness and the preplasma extent. The simulations run for $400 \kern0.2em \mathrm{fs}$ over which the initial ${\sim} \kern0.1em 200 \kern0.2em \mathrm{fs}$ are for the pulse to travel from the simulation boundary to the interaction point, at the $(0,0)$ coordinates; the initial $160 \kern0.2em \mathrm{fs}$ allow only electromagnetic fields evolution in order to reduce the computational time. A time-step multiplier factor of $0.8$ is set, resulting in $21199$ steps with a time-step of ${\sim} \kern0.1em 19 \kern0.2em \mathrm{as}$ (or ${\sim} \kern0.1em 1.77$ times the cell size over the speed of light).

\par The simulation boundaries are set to ``open'' for both fields and particles in both directions (allowing them to exit the simulation), except for one of the x-boundaries which is set to ``simple-laser'' allowing the imposing of an electromagnetic pulse source. However, due to the relatively large size of the simulation box, neither particles nor fields reach box the boundary at the end of the simulation. The laser pulse is launched at an angle of $15 \degree$ with respect to the $\bf{\hat{x}}$ axis, according to the equations governing the propagation of a focused laser beam \cite{Siegman1986}. The beam is characterised by a Gaussian profile in both temporal and spatial directions, with a temporal FWHM of $30 \kern0.2em \mathrm{fs}$ and a spatial FWHM of $5 \kern0.2em \mathrm{\upmu m}$. An offset of two standard deviations is added to the temporal profile of the pulse in order to allow for the rising pulse part to be sufficiently generated, while an offset is also added to the spatial direction allowing the pulse to focus on the $(0,0)$ coordinate. The laser wavelength is set to $810 \kern0.2em \mathrm{nm}$ and the pulse is P-polarised.  In this work the results of three groups of simulations are presented according to the pulse peak intensity, having values of $1.85 \kern0.1em {\times} \kern0.1em 10^{20} \kern0.2em \mathrm{W cm^{-2}}$, $1.85 \kern0.1em {\times} \kern0.1em 10^{21} \kern0.2em \mathrm{W cm^{-2}}$ and $1.85 \kern0.1em {\times} \kern0.1em 10^{22} \kern0.2em \mathrm{W cm^{-2}}$.

\par Four particle species are used in the code (including electrons, protons, carbon ions ($\mathrm{C}$) and oxygen ions ($\mathrm{O}$) composing a biaxially-oriented polyethylene terephthalate \cite{DeMeuse2011} foil (commonly known as Mylar), represented by a chemical formula of $\mathrm{H_8 C_{10} O_4}$. A charge of $-1, 1, 6$ and $8$ times the elementary charge and a mass of $1, 1836.15, 12 \kern0.1em {\times} \kern0.1em 1836.15$ and $16 \kern0.1em {\times} \kern0.1em 1836.15$ times the electron rest mass is set for electrons, protons, $\mathrm{C}$ and $\mathrm{O}$ respectively. The number density of each specie is imported into the simulation from four ``.dat'' files containing the number density as calculated by the hydrodynamic simulations, described in Sec. \ref{hydro} and Sec. \ref{altered}.


\section{PIC Simulations Results} \label{Results}

\par As it will become obvious from the material presented in the current section, the existence of a preformed preplasma at a foil target significantly affects the laser-target interaction. Therefore, the performed PIC simulations include the density distributions obtained from hydrodynamic simulations. In this section the results of a multi-parametric study of the preplasma effect on the laser-foil interaction is presented, emphasising on the resulting proton acceleration. Different preplasma distributions are considered for Mylar foils, resulting due to different combinations of the ASE pedestal intensity ($1.6 \kern0.1em {\times} \kern0.1em 10^9 \kern0.2em \mathrm{W cm^{-2}}$ and $1.6 \kern0.1em {\times} \kern0.1em 10^{10} \kern0.2em \mathrm{W cm^{-2}}$) and foil thickness ($0.1 \kern0.2em \mathrm{\upmu m}$, $0.3 \kern0.2em \mathrm{\upmu m}$ and $0.9 \kern0.2em \mathrm{\upmu m}$). Furthermore, the effect of the target thickness on proton acceleration is examined for main pulse intensities of  $1.85 \kern0.1em {\times} \kern0.1em 10^{20} \kern0.2em \mathrm{W cm^{-2}}$, $1.85 \kern0.1em {\times} \kern0.1em 10^{21} \kern0.2em \mathrm{W cm^{-2}}$ and $1.85 \kern0.1em {\times} \kern0.1em 10^{22} \kern0.2em \mathrm{W cm^{-2}}$, revealing that the foil optimal thickness strongly depends on the preplasma parameters.

\begin{figure}[!hb]
  \includegraphics[width=\linewidth]{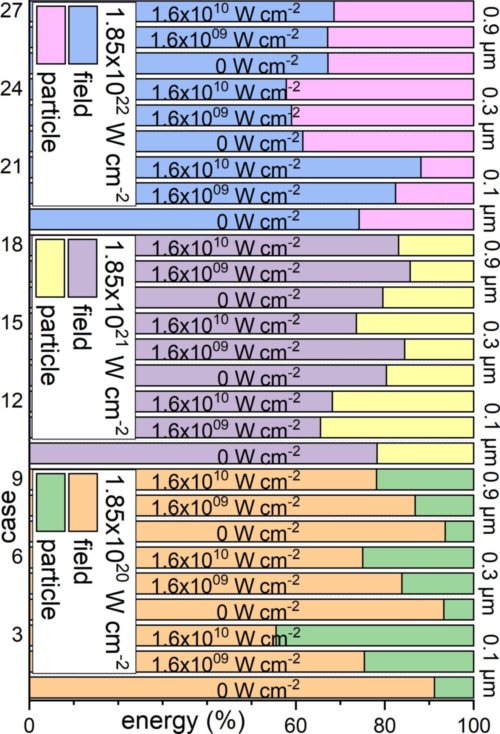}
  \caption{Percentages of the energy absorbed by the target versus the pulse energy reflected/transmitted. The orange/green, purple/yellow and blue/pink sections correspond to main pulse intensities of $1.85 \kern0.1em {\times} \kern0.1em 10^{20} \kern0.2em \mathrm{W cm^{-2}}$, $1.85 \kern0.1em {\times} \kern0.1em 10^{21} \kern0.2em \mathrm{W cm^{-2}}$ and $1.85 \kern0.1em {\times} \kern0.1em 10^{22} \kern0.2em \mathrm{W cm^{-2}}$. Each section corresponds to cases of different preplasmas, defined by a combination of the foil thickness and ASE pedestal intensity, as noted on the figure for each case.}
  \label{fig:pulse_particle}
\end{figure}


\subsection{Energy Absorption} \label{absorption}

\par As explained in detail in Sec. \ref{characterisation}, in our PIC simulations a laser pulse is launched at an angle of $15 \degree$ with the target normal axis. The simulations are set in such a way that if the foil has a steep density profile then the laser-foil interaction takes place at the location $(0,0)$. However, the existence of a preplasma density gradient in combination with the target dislocation due to the ASE pedestal shifts the location of the interaction point, as can be realised from Fig. \ref{fig:preplasma_lineout_15deg}. For simplicity, let us consider the simulated case corresponding to the least target dislocation. as shown by the dashed blue line in Fig. \ref{fig:preplasma_lineout_15deg} ($0.9 \kern0.2em \mathrm{\upmu m}$ thick foil with an ASE pedestal intensity of $1.6 \kern0.1em {\times} \kern0.1em 10^9 \kern0.2em \mathrm{W cm^{-2}}$).

\begin{figure}[!b]
  \includegraphics[width=\linewidth]{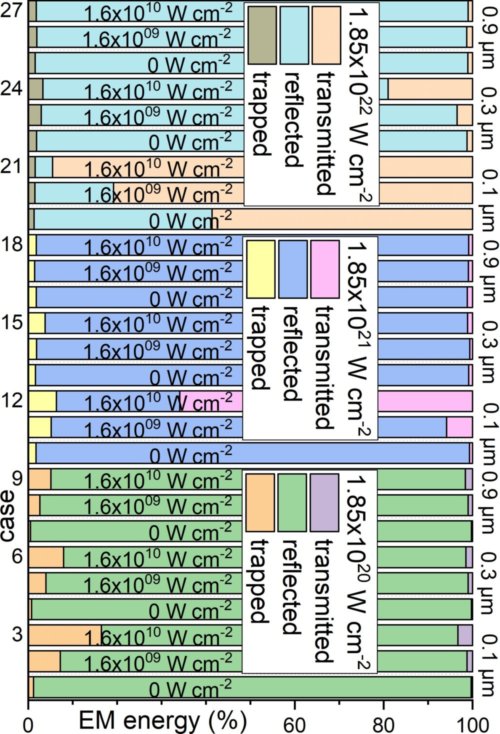}
  \caption{Percentages of the transmitted, reflected and captured electromagnetic energy.The orange/green/purple, yellow/blue/pink and khaki/cyan/melon sections correspond to main pulse intensities of $1.85 \kern0.1em {\times} \kern0.1em 10^{20} \kern0.2em \mathrm{W cm^{-2}}$, $1.85 \kern0.1em {\times} \kern0.1em 10^{21} \kern0.2em \mathrm{W cm^{-2}}$ and $1.85 \kern0.1em {\times} \kern0.1em 10^{22} \kern0.2em \mathrm{W cm^{-2}}$. Each section corresponds to cases of different preplasmas, defined by a combination of the foil thickness and ASE pedestal intensity, as noted on the figure for each case.}
  \label{fig:reflected_transmitted}
\end{figure}

\par As the main pulse travels towards the interaction point, a long preplasma region is located in front of the foil front surface. However, the number density of this preplasma is well below the critical number density and the pulse continues propagating without disruption. Only at a sub-micron distance from the peak number density the pulse reaches the contour of the critical density. However, due to the relativistic correction of the critical number density the interaction location is shifted even closer to the location of the maximum target number density (approximately equal the density of a flat-top profile). Since the relativistically corrected critical number density depends on the electric field amplitude, the rising profile of the pulse faces a surface at a much lower density, with the limit given by the classical critical number density (baseline of Fig. \ref{fig:preplasma_lineout_15deg}).

\par As a consequence, the pulse interacts with the target in a preplasma region defined by the classical and critical number densities. In that region the laser pulse deposits a fraction of its' energy to hot electrons while the rest is reflected/transmitted towards free space. The percentage of radiation absorbed by the target strongly depends on the preplasma density, as can be seen in Fig. \ref{fig:pulse_particle}. The figure summarises the amount of energy transferred by the pulse to the particles for all cases presented in this work. A general conclusion that can be extracted from Fig. \ref{fig:pulse_particle} is that for lower main pulse intensities ($1.85 \kern0.1em {\times} 10^{20} \kern0.2em \mathrm{W cm^{-2}}$ - orange/green colour in Fig. \ref{fig:pulse_particle}) a higher preplasma formation (ASE pedestal intensity $1.6 \kern0.1em {\times} \kern0.1em 10^{10} \kern0.2em \mathrm{W cm^{-2}}$) benefits the energy transfer towards particles, while as the main pulse intensity is increased ($1.85 \kern0.1em {\times} \kern0.1em 10^{22} \kern0.2em \mathrm{W cm^{-2}}$ - blue/pink colour in Fig. \ref{fig:pulse_particle}), this beneficial tendency ceases. The significantly lower energy transfer observed for thinner targets at higher intensities can be explained by the significantly higher amount of the main pulse passing through the foil, as can be seen in Fig. \ref{fig:B_field} and explained by Fig. \ref{fig:evolution}.


\subsection{Reflected, Transmitted and Trapped Fields} \label{field}

\begin{figure}[!t] 
  \includegraphics[width=\linewidth]{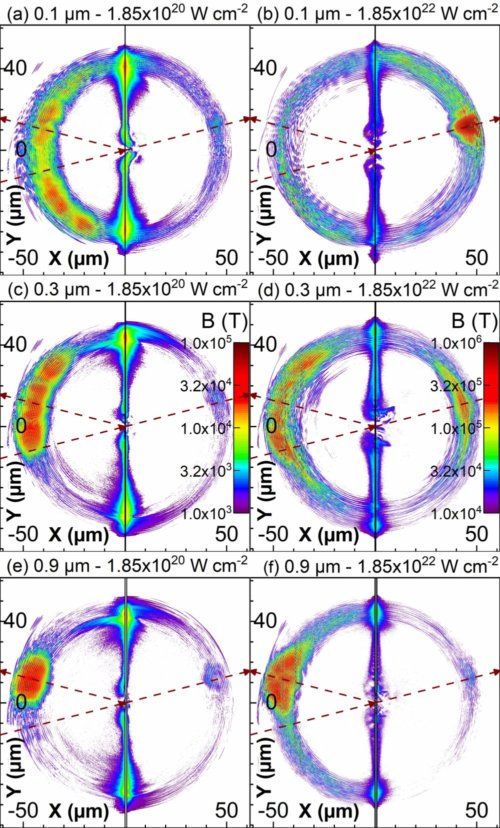}
  \caption{Magnetic field corresponding to an ASE pedestal intensity of $1.6 \kern0.1em {\times} \kern0.1em 10^{10} \kern0.2em \mathrm{W cm^{-2}}$ at a simulation time of $400 \kern0.2em \mathrm{fs}$, for various foil thicknesses as labelled on the figure. The left column corresponds to a main pulse intensity of $1.85 \kern0.1em {\times} \kern0.1em 10^{20}\kern0.2em \mathrm{W cm^{-2}}$ and the right column to $1.85 \kern0.1em {\times} \kern0.1em 10^{22} \kern0.2em \mathrm{W cm^{-2}}$.}
  \label{fig:B_field}
\end{figure}

\par Before considering the electron/ion spectra it is crucial to examine the laser pulse behaviour before and after the laser-foil interaction takes place. Since a fraction of the laser pulse energy is transferred into the particles, the electromagnetic energy after the interaction is less compared to the initial, as seen in Fig. \ref{fig:pulse_particle}. However, in order to compare the various simulated cases, the electromagnetic energy after the laser-foil interaction is scaled to $100 \kern0.2em \%$ (besides, an insignificant variation of ${ {\sim} \kern0.1em \pm 15 \kern0.2em \% }$ exists between the simulated cases).

\par The behaviour of the pulse following the laser-foil interaction is shown schematically in Fig. \ref{fig:B_field}, where the magnetic field amplitude is shown in a contour form at a time of ${ {\sim} \kern0.1em 200 \kern0.2em \mathrm{fs} }$ after the interaction. The new spatial field distribution can be separated in three categories, which represent a reflected, a transmitted and a trapped field by the target. As explained in Sec. \ref{absorption}, the pulse is reflected at a region defined by the classical and relativistically corrected critical number densities. For the case where no preplasma exists on the target, the reflected part of the electromagnetic energy is ${\sim} \kern0.1em 99 \kern0.2em \%$ for a main pulse intensity of $1.85 \kern0.1em {\times} \kern0.1em 10^{20} \kern0.2em \mathrm{W cm^{-2}}$, as seen in Fig. \ref{fig:reflected_transmitted}.

\par An interesting behaviour of the pulse reflection due to the target modified surface can be seen in the left column of Fig \ref{fig:B_field}, where the target survives the interaction with the main pulse (in contrast with the right side of the diagram where a hole is created in thinner foils). There, the reflected pulse by a $0.9 \kern0.2em \mathrm{\upmu m}$ thick foil appears highly structured and with a similar shape to the incoming pulse. On the other hand, for a $0.1 \kern0.2em \mathrm{\upmu m}$ thick target the pulse is reflected in all radial directions. This behaviour can be realised by the strongly modified target surface due to the ASE pedestal, as seen in Fig. \ref{fig:preplasma}-(b). Therefore, the main pulse does not interact with a relatively flat target as in the case of the $0.9 \kern0.2em \mathrm{\upmu m}$ thick target, but rather with a complex curved surface, where the curvature varies for different electron density contours.

\par  In most cases, the transmitted pulse is ${\sim} \kern0.1em 1 \kern0.2em \%$, which can be explained in terms of relativistic transparency of the target \cite{Esirkepov2014, Vshivkov1998}. However, for thin ($0.1 \kern0.2em \mathrm{\upmu m}$) targets and higher main pulse intensities ($1.85 \kern0.1em {\times} \kern0.1em 10^{21} \kern0.2em \mathrm{W cm^{-2}}$ and $1.85 \kern0.1em {\times} \kern0.1em 10^{22} \kern0.2em \mathrm{W cm^{-2}}$) the transmitted electromagnetic energy appears to be significantly higher than the reflected. The explanation of this behaviour is related to the preplasma formation. As seen by the top dashed line in Fig. \ref{fig:preplasma_lineout_15deg}, for a $0.1 \kern0.2em \mathrm{\upmu m}$ thick target and an ASE pedestal intensity of $1.6 \kern0.1em {\times} \kern0.1em 10^{10} \kern0.2em \mathrm{W cm^{-2}}$, the target is opaque since the target highest number density is lower than the relativistically corrected critical number density. Therefore, most of the incident pulse is allowed to pass through the target, as shown in Fig. \ref{fig:B_field}-(b). A similar explanation applies for an intensity of $1.85 \kern0.1em {\times} \kern0.1em 10^{21} \kern0.2em \mathrm{W cm^{-2}}$, since for a $0.1 \kern0.2em \mathrm{\upmu m}$ thick foil the maximum number target density is marginally above the relativistically corrected critical number density. As the main pulse interacts with the target, this small difference in densities is reversed, allowing a large fraction of the pulse to pass through, as seen in Fig. \ref{fig:B_field}-(d).

\par A third portion of the pulse is trapped in the foil, initially in the laser-foil interaction point and then propagating outwards. This effect has been seen experimentally in various instances \cite{Borghesi2002, Borghesi2003, Quinn2009, Tokita2015} in the form of a positively charged propagating pulse. In our simulations the trapped field is followed by an electron population, as can be seen by the $\left( y , p_y \right)$ diagram in Fig. \ref{fig:PSe20} and Fig. \ref{fig:PSe22}; there, $p_y$ forms a clear peak in the vicinity of the trapped (by the foil) portion of the laser pulse. Since $p_x$ is much smaller (appears as two tiny brown dots in the $\left( x , p_y \right)$ diagram, at the same location where $\left( y , p_y \right)$ peaks) than $p_y$, those electrons following the trapped pulse move almost parallel to the foil surface. Those electrons following the tail of the pulse are also described in Refs. \onlinecite{Tokita2011, Tokita2015}, where Ref. \onlinecite{Tokita2015} also contains an extensive description of the pulse.


\subsection{Electron Spectra} \label{Electron Spectra}

\begin{figure}[!t] 
  \includegraphics[width=\linewidth]{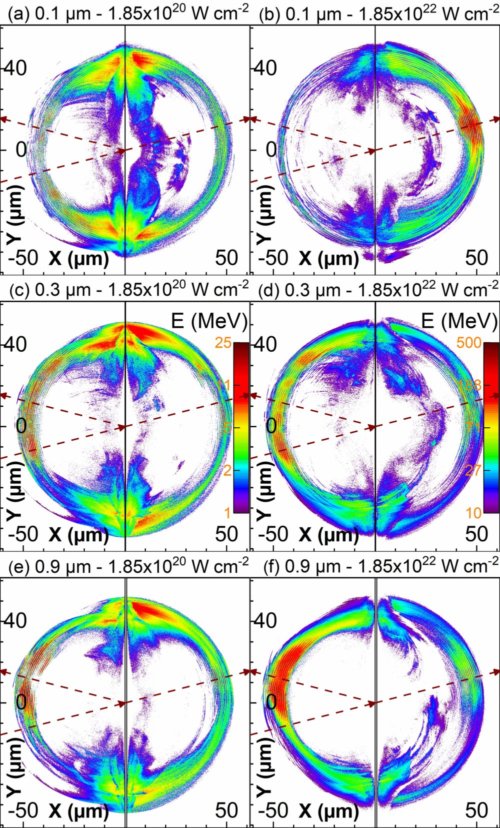}
  \caption{Electron kinetic energy for a preplasma corresponding to an ASE pedestal intensity of $1.6 \kern0.1em {\times} \kern0.1em 10^{10} \kern0.2em \mathrm{W cm^{-2}}$ at a simulation time of $400 \kern0.2em \mathrm{fs}$, for various foil thicknesses as labelled on the figure. The left column corresponds to a main pulse intensity of $1.85 \kern0.1em {\times} \kern0.1em 10^{20} \kern0.2em \mathrm{W cm^{-2}}$ and the right column to $1.85 \kern0.1em {\times} \kern0.1em 10^{22} \kern0.2em \mathrm{W cm^{-2}}$.}
  \label{fig:Ekbar_e}
\end{figure}

\par Once the laser pulse reaches the target surface it transfers part of its energy to electrons creating a population of hot electrons. These hot electrons are then affected by the laser pulse. By the end of the interaction (${\sim} \kern0.1em 200 \mathrm{fs}$), several distinct electron groups are formed, as can be visually seen in Fig. \ref{fig:Ekbar_e}; the contour figure represents the average kinetic energy of electrons at the end of the simulation for a collection of foil thicknesses, with a main pulse intensity of $1.85 \kern0.1em {\times} \kern0.1em 10^{20} \kern0.2em \mathrm{W cm^{-2}}$ shown on the left column and an intensity of $1.85 \kern0.1em {\times} \kern0.1em 10^{22} \kern0.2em \mathrm{W cm^{-2}}$ on the right column, where in all cases the ASE pedestal intensity is $1.6 \kern0.1em {\times} \kern0.1em 10^{10} \kern0.2em \mathrm{W cm^{-2}}$. Interestingly, the overall pattern of Fig. \ref{fig:Ekbar_e} represents significant similarities to the pattern of the magnetic field distribution, as shown in Fig. \ref{fig:B_field}.

\begin{figure}[!hb]
  \includegraphics[width=\linewidth]{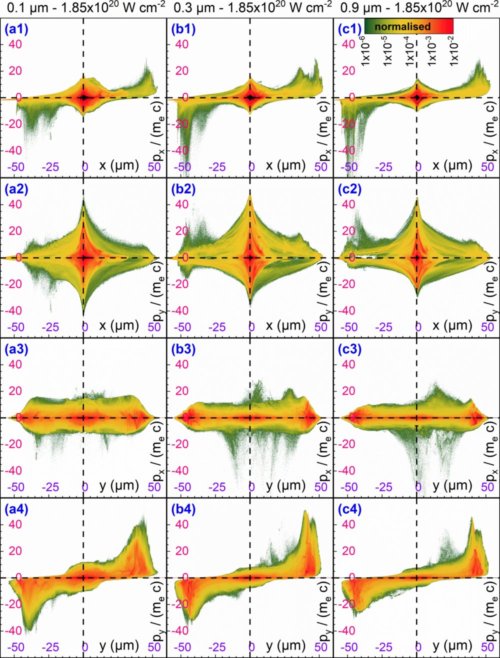}
  \caption{Electron $\left( x , p_x \right)$ (1st row), $\left( x , p_y \right)$ (2nd row), $\left( y , p_x \right)$ (3rd row) and $\left( y , p_y \right)$ (4th row) diagrams corresponding to $0.1 \kern0.2em \mathrm{\upmu m}$ (1st column), $0.3 \kern0.2em \mathrm{\upmu m}$ (2nd column) and $0.9 \kern0.2em \mathrm{\upmu m}$ (3rd column) thick foils respectively, for an ASE pedestal intensity of $1.6 \kern0.1em {\times} \kern0.1em 10^{10} \kern0.2em \mathrm{W cm^{-2}}$ and a main pulse intensity of $1.85 \kern0.1em {\times} \kern0.1em 10^{20} \kern0.2em \mathrm{W cm^{-2}}$ at a simulation time of $400 \kern0.2em \mathrm{fs}$.}
  \label{fig:PSe20}
\end{figure}

\begin{figure}[!hb]
  \includegraphics[width=\linewidth]{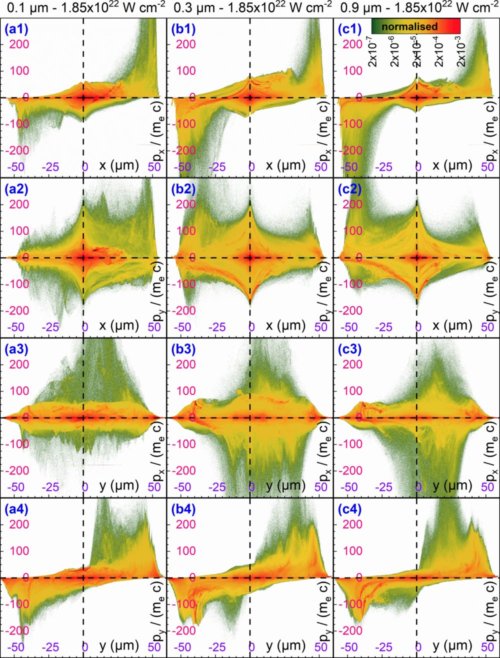}
  \caption{Electron $\left( x , p_x \right)$ (1st row), $\left( x , p_y \right)$ (2nd row), $\left( y , p_x \right)$ (3rd row) and $\left( y , p_y \right)$ (4th row) diagrams corresponding to $0.1 \kern0.2em \mathrm{\upmu m}$ (1st column), $0.3 \kern0.2em \mathrm{\upmu m}$ (2nd column) and $0.9 \kern0.2em \mathrm{\upmu m}$ (3rd column) thick foils respectively, for an ASE pedestal intensity of $1.6 \kern0.1em {\times} \kern0.1em 10^{10} \kern0.2em \mathrm{W cm^{-2}}$ and a main pulse intensity of $1.85 \kern0.1em {\times} \kern0.1em 10^{22} \kern0.2em \mathrm{W cm^{-2}}$ at a simulation time of $400 \kern0.2em \mathrm{fs}$.}
  \label{fig:PSe22}
\end{figure}


\begin{figure}[!ht]
  \includegraphics[width=\linewidth]{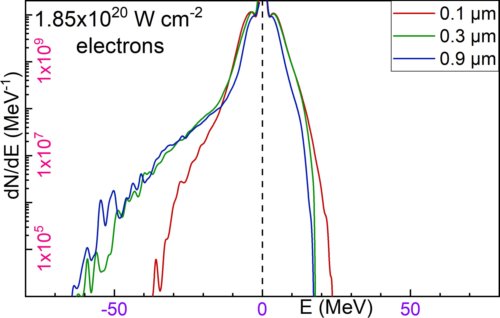}
  \caption{Electron spectra corresponding to $0.1 \kern0.2em \mathrm{\upmu m}$ (red line), $0.3 \kern0.2em \mathrm{\upmu m}$ (green line) and $0.9 \kern0.2em \mathrm{\upmu m}$ (blue line) for an ASE pedestal intensity of $1.6 \kern0.1em {\times} \kern0.1em 10^{10} \kern0.2em \mathrm{W cm^{-2}}$ and a main pulse intensity of $1.85 \kern0.1em {\times} \kern0.1em 10^{20} \kern0.2em \mathrm{W cm^{-2}}$. The ``negative'' energy axis denotes electrons located at the target front region, while the positive energy axis denotes electrons at the target rear region.}
  \label{fig:spectrum_eE20}
\end{figure}


\begin{figure}[!hb]
  \includegraphics[width=\linewidth]{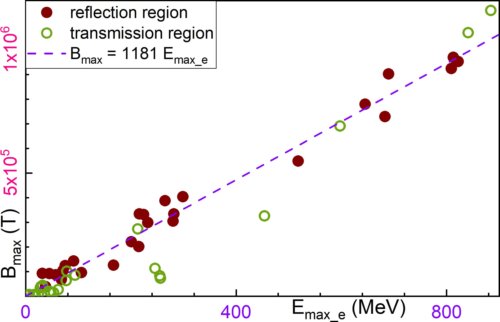}
  \caption{The maximum magnetic field versus the maximum electron energy. The solid brown circles correspond to the region of the reflected laser pulse, while the hollow green circles to the region of the transmitted pulse.The purple dashed line is a linear fit to the data corresponding to the reflected pulse region. No trend-line can be plotted for the transmitted energy due to diffraction effects in the cases where the target is destroyed by the pulse.}
  \label{fig:e_scale}
\end{figure}


\begin{figure}[!ht]
  \includegraphics[width=\linewidth]{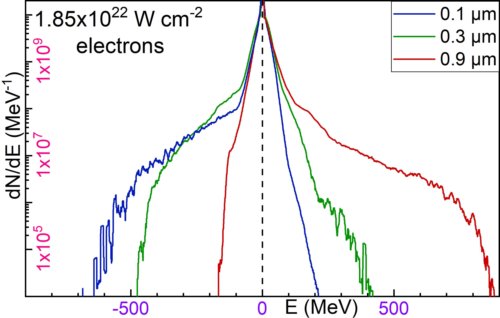}
  \caption{Electron spectra corresponding to $0.1 \kern0.2em \mathrm{\upmu m}$ (red line), $0.3 \kern0.2em \mathrm{\upmu m}$ (green line) and $0.9 \kern0.2em \mathrm{\upmu m}$ (blue line) for an ASE pedestal intensity of $1.6 \kern0.1em {\times} \kern0.1em 10^{10} \kern0.2em \mathrm{W cm^{-2}}$ and a main pulse intensity of $1.85 \kern0.1em {\times} \kern0.1em 10^{22} \kern0.2em \mathrm{W cm^{-2}}$. The ``negative'' energy axis denotes electrons located at the target front region, while the positive energy axis denotes electrons at the target rear region.}
  \label{fig:spectrum_eE22}
\end{figure}

\par One would not expect a correlation of the laser pulse and electron distribution due to the expel of electrons from the pulse region due to reflection at the ponderomotive potential barrier. However, as it is shown in Refs. \onlinecite{Wang1998, Zhy1998, Wang2001, Braenzel2017}, under some conditions the laser pulse can capture an electron population, significantly increasing its' energy. One necessary condition for this to happen, is that the pulse intensity must exceed $10^{22} \kern0.2em \mathrm{W cm^{-2}}$, which corresponds to the case shown on the right column of Fig. \ref{fig:B_field} and Fig. \ref{fig:Ekbar_e}. Although our simulations show that the electron capture can happen even for the pulse intensity of $10^{20} \kern0.2em \mathrm{W cm^{-2}}$ the effect is not so prominent as in the case of higher intensity, where the a clearly localised high energy electron population is present (see the transmitted pulse region for the $0.1 \kern0.2em \mathrm{\upmu m}$ thick foil and the reflected pulse region for the $0.9 \kern0.2em \mathrm{\upmu m}$ foil). The second condition of the electrons to be captured is that their propagation axis must be near the propagation axis of the pulse; since the electrons are initially expelled in all directions from the target, it is evident that this condition is unavoidably met by some fraction of the initial hot electron population.

\par The physical explanation of the electron captured is given in Ref. \onlinecite{Wang2001}, indicating that since the pulse phase velocity near the focal region is less than $c$, some electrons can be kept in phase with the pulse for long times, gaining a considerable amount of energy. The dependency of the maximum electron energy on the field amplitude can be seen in Fig. \ref{fig:e_scale}, where a linear fit can be well applied. The solid circles correspond to the target front surface, while the hollow circles to the target rear. As expected, the points corresponding to the target rear do not initially follow a linear trend; however, for thinner targets and higher main pulse intensities the target is destroyed allowing the field to pass through, explaining why the electrons in these cases follow the linear fit.

\par The creation of a localised electron population due to the laser-foil interaction has also been demonstrated experimentally, as in Refs. \onlinecite{Tian2012, Thevenet2015}. Their results demonstrate a clearly localised electron signal with an approximately circular profile; in both works, a small  deep is observed in one side of the signal, near the maximum. This experimental observation can potentially explain the small asymmetry of the mean electron energy in the region of the reflected pulse, as seen by the electron mean energy distribution along the dashed reflection line in Fig. \ref{fig:Ekbar_e}(e-f). In contrast, no asymmetry is observed for the reflected field distribution.

\par Another distinct electron population corresponds to the fraction of the pulse that is captured by the foil surface. These electrons are extremely localised, as can be seen by the $\left( y , p_y \right)$ diagrams in Fig. \ref{fig:PSe20} and Fig. \ref{fig:PSe22}, where they appear as two distinct peaks located in the vicinity of the captured pulse. Although their energy is not as high as in the case of the reflected pulse captured electrons, it is still in the multi-$\mathrm{MeV}$ region. Their existence and properties have also been observed in Refs. \onlinecite{Tokita2011, Tokita2015}.

\par The electron spectrum for an ASE pedestal intensity of $1.6 \kern0.1em {\times} \kern0.1em 10^{10} \kern0.2em \mathrm{W cm^{-2}}$ and a main pulse intensity of $1.85 \kern0.1em {\times} \kern0.1em 10^{20} \kern0.2em \mathrm{W cm^{-2}}$ is shown in Fig. \ref{fig:spectrum_eE20}, while for a main pulse intensity of $1.85 \kern0.1em {\times} \kern0.1em 10^{22} \kern0.2em \mathrm{W cm^{-2}}$ it is shown in Fig. \ref{fig:spectrum_eE22}; in both figures, the $0.1 \kern0.2em \mathrm{\upmu m}$, $0.3 \kern0.2em \mathrm{\upmu m}$ and $0.9 \kern0.2em \mathrm{\upmu m}$ thick foils are indicated by a red, green and blue line respectively. The ``negative'' energy axis represents the spectrum for electrons located on the target front area, while the positive energy axis correspond to electrons on the target rear area. In order to reduce the noise from the spectrum, electrons located on the initial foil location or near the laser-foil interaction region (lower energy hot electron cloud that moves both inwards and outwards the target) are excluded from the spectrum. These excluded electrons appear as electrons refluxing through the target in the form of a hot electron cloud, as can be seen by the inner population (near the foil target) presented in Fig. \ref{fig:Ekbar_e}. It was demonstrated experimentally \cite{Yogo2017} that by controlling their distribution one can enhance ion acceleration, while a theoretical description of the stochastic electron heating is given in Ref. \onlinecite{Bulanov2015}.

\par Both Fig. \ref{fig:spectrum_eE20} and Fig. \ref{fig:spectrum_eE22} exhibit a very similar trend, for both the electron spectra from the target front and rear. Each spectral line is a superposition of three fundamental populations. The first group is a low energy electron noise, in the ${\sim} \kern0.1em 1 \kern0.2em \mathrm{MeV}$ range. The second group is a Maxwellian-like distribution which for Fig. \ref{fig:spectrum_eE20} and Fig. \ref{fig:spectrum_eE22} is in the range $< 15 \kern0.2em \mathrm{MeV}$ and $< 150 \kern0.2em \mathrm{MeV}$ respectively; due to the low energy noise population, the rising profile is not clearly resolved in Fig. \ref{fig:spectrum_eE22}. This population corresponds to the electrons travelling along the captured portion of the pulse; therefore, it is symmetric for both the electron spectra from the target front and rear.

\par The third group is an exponentially decaying distribution, characterised by a sharp cut-off. No fitting is presented for these distributions due to their 2D nature, although the temperature is in the $\mathrm{MeV}$ range. These hot electrons correspond to a clearly detached electron population, seen schematically in Fig. \ref{fig:Ekbar_e}; it is also realised as the distinct peaks formed on the $\left( x , p_x \right)$ diagram, shown in Fig. \ref{fig:PSe20} and Fig. \ref{fig:PSe22}. For $1.85 \kern0.1em {\times} \kern0.1em 10^{20} \kern0.2em \mathrm{W cm^{-2}}$ main pulse intensity, the electron spectrum from the target front region reaches ${\sim} \kern0.1em 60 \kern0.2em \mathrm{MeV}$ for a $0.9 \kern0.2em \mathrm{\upmu m}$ thick target and ${\sim} \kern0.1em 40 \kern0.2em \mathrm{MeV}$ for $0.1 \kern0.2em \mathrm{\upmu m}$. If one also considers that the thin target case has a much higher particle to field ratio (see Fig. \ref{fig:pulse_particle}) it becomes evident that thin targets will give a significantly higher proton energy. For a main pulse intensity of $1.85 \kern0.1em {\times} \kern0.1em 10^{22} \kern0.2em \mathrm{W cm^{-2}}$ the picture is similar, where the $0.9 \kern0.2em \mathrm{\upmu m}$ thick target gives an electron energy of ${\sim} \kern0.1em 600 \kern0.2em \mathrm{MeV}$ on the target front surface region. However, at such an intensity the central part of the $0.1 \kern0.2em \mathrm{\upmu m}$ thick target is transparent to the pulse (see fig. \ref{fig:preplasma_lineout_15deg}), driving the captured electrons at an energy of ${\sim} \kern0.1em 800 \kern0.2em \mathrm{MeV}$, located on the target rear region.


\subsection{Proton Spectra} \label{Proton Spectra}

\begin{figure}[!hb]
  \includegraphics[width=\linewidth]{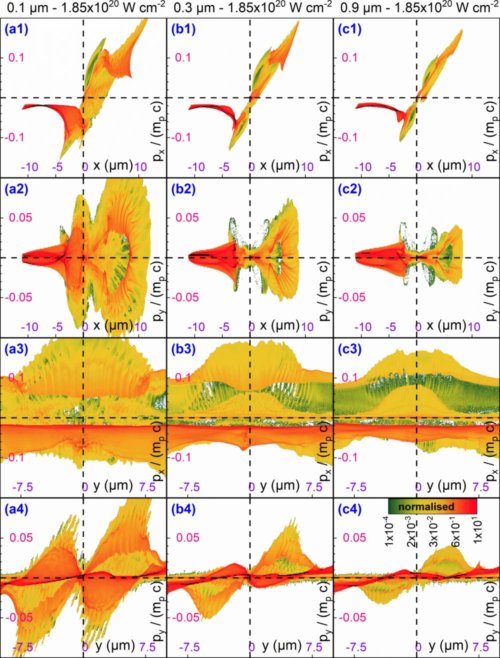}
  \caption{Proton $\left( x , p_x \right)$ (1st row), $\left( x , p_y \right)$ (2nd row), $\left( y , p_x \right)$ (3rd row) and $\left( y , p_y \right)$ (4th row) diagrams corresponding to $0.1 \kern0.2em \mathrm{\upmu m}$ (1st column), $0.3 \kern0.2em \mathrm{\upmu m}$ (2nd column) and $0.9 \kern0.2em \mathrm{\upmu m}$ (3rd column) thick Mylar foils for an ASE pedestal intensity of $1.6 \kern0.1em {\times} \kern0.1em 10^{10} \kern0.2em \mathrm{W cm^{-2}}$ and a main pulse intensity of $1.85 \kern0.1em {\times} \kern0.1em 10^{22} \kern0.2em \mathrm{W cm^{-2}}$.}
  \label{fig:PSh20}
\end{figure}

\par As a direct consequence of the hot electron generation and ejection from the target is the establishment of an electric field normal to the target surface (target normal sheath acceleration (TNSA) model \cite{Macchi2013, Klimo2008}), which accelerates protons and ions, mainly from the target rear surface towards the vacuum region. This acceleration mechanism is dominant at intensities of $10^{20} \kern0.2em \mathrm{W cm^{-2}}$ (first intensity group in this work). However, at higher intensities (${\sim} \kern0.1em 10^{22} \kern0.2em \mathrm{W cm^{-2}}$) and by setting a circular polarisation to the laser pulse, the radiation pressure acceleration (RPA) mechanism \cite{Robinson2008, Bulanov2016} becomes important. There, the pressure associated with the laser pulse can accelerate the whole foil (located in the interaction region). Therefore, the foil thickness is related with how efficient RPA is, since thicker foils are harder to be accelerated as a unity. However, one disadvantage of the RPA mechanism is that a very thin foil can be destroyed for a higher intensity pulse and then the proton acceleration is significantly reduced. In our work, the laser polarisation is kept linear since RPA is not examined; however, the effect of increased pulse intensity (linearly polarised) and foil thickness reduction are presented. In a different scenario, the Coulomb Explosion \cite{Bulanov2016} (CE) acceleration mechanism could dominate. The requirements of this mechanism to be efficient is a high amplitude high contrast laser pulse incident on a thin foil target. Then, if all electrons present in the focal spot region are able to escape the target, the foil is left only with ions during a time period greater than the electron ejection time and smaller than the proton response time. As a result of that, the ions will feel strong repulsive forces leading to the CE. The resulting ion spectrum is characterised by a non-thermal distribution with a maximum energy determined by the maximum electrostatic potential created in the ion cluster prior the CE.

\par A vital understanding of the behaviour of the accelerated protons can be realised by the space-momentum diagrams, as shown in Fig. \ref{fig:PSh20} which corresponds to an ASE pedestal intensity of  $1.6 \kern0.1em {\times} \kern0.1em 10^{10} \kern0.2em \mathrm{W cm^{-2}}$ and a main pulse intensity of  $1.85 \kern0.1em {\times} \kern0.1em 10^{20} \kern0.2em \mathrm{W cm^{-2}}$. The figure is also organised in three columns, according to different foil thicknesses. The first important observation comes from both the $\left( x , p_x \right)$ and $\left( y , p_x \right)$ diagrams, where it becomes obvious that the protons on the rear foil region can be separated in two groups, according to their energy. The time-evolution of the diagrams (not presented here) reveals a different origin of these two populations. The higher energy group originates from the target rear surface, in agreement with the TNSA theory; the second group initiates from the target front surface, it then penetrates the target and reappears from the rear surface as the lower energy group. As time evolves, the front of the lower energy group reaches the tail of the higher energy group and they then appear as a continuous distribution in space. A similar behaviour has also been observed in Ref. \onlinecite{Nakamura2003}. Although a third group can be considered as also initiating from the target front surface and moving towards the opposite direction (compared to the higher energy protons), due to their significantly lower energy they can be ignored. Actually, the partial symmetry of the $\left( x , p_x \right)$ diagram indicates that the protons initiating from the target front surface and then travel towards opposite directions have comparable energies, although much different divergence.

\par
\begin{figure}[!ht]
  \includegraphics[width=\linewidth]{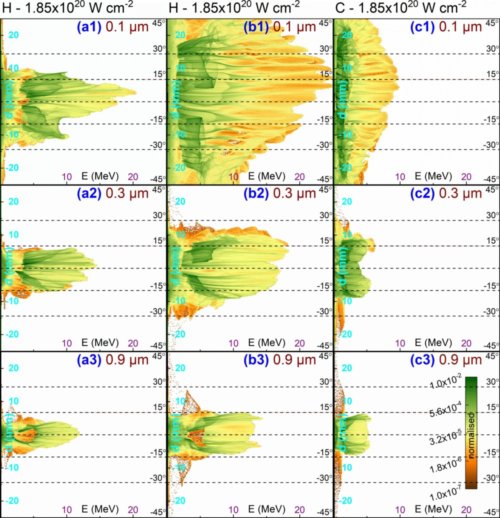}
  \caption{Projection of ion trajectories on the plane located at a distance of $25 \kern0.1em \mathrm{mm}$ from the front target surface. All contours correspond to a main pulse intensity of $1.85 \kern0.1em {\times} \kern0.1em 10^{20} \kern0.2em \mathrm{W cm^{-2}}$ and a simulation time of $400 \kern0.2em \mathrm{fs}$, while the first, second and third row correspond to an initial foil thickness of $0.1 \kern0.2em \mathrm{\upmu m}$, $0.3 \kern0.2em \mathrm{\upmu m}$ and $0.9 \kern0.2em \mathrm{\upmu m}$ respectively. The first column corresponds to an ASE pedestal of $1.6 \kern0.1em {\times} \kern0.1em 10^{9} \kern0.2em \mathrm{W cm^{-2}}$ and the second column to $1.6 \kern0.1em {\times} \kern0.1em 10^{10} \kern0.2em \mathrm{W cm^{-2}}$, both for hydrogen. The third column corresponds to an ASE pedestal of $1.6 \kern0.1em {\times} \kern0.1em 10^{10} \kern0.2em \mathrm{W cm^{-2}}$ and carbon ions. The particle detection plane extends to $\pm 25 \kern0.2em \mathrm{mm}$, corresponding to a maximum divergence angle of $45 \degree$.}
  \label{fig:RCF20}
\end{figure}

\begin{figure}[!ht]
  \includegraphics[width=\linewidth]{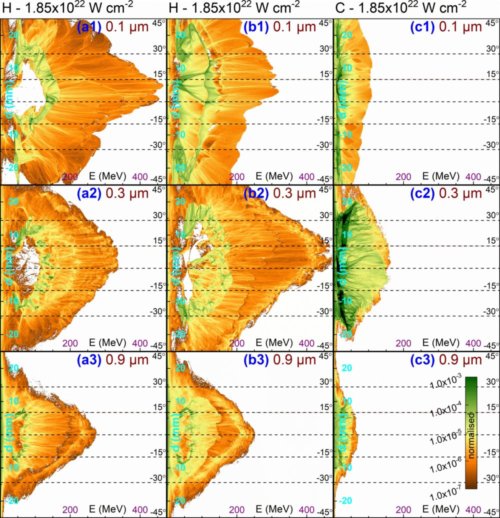}
  \caption{Projection of ion trajectories on the plane located at a distance of $25 \kern0.1em \mathrm{mm}$ from the front target surface. All contours correspond to a main pulse intensity of $1.85 \kern0.1em {\times} \kern0.1em 10^{22} \kern0.2em \mathrm{W cm^{-2}}$ and a simulation time of $400 \kern0.2em \mathrm{fs}$, while the first, second and third row correspond to an initial foil thickness of $0.1 \kern0.2em \mathrm{\upmu m}$, $0.3 \kern0.2em \mathrm{\upmu m}$ and $0.9 \kern0.2em \mathrm{\upmu m}$ respectively. The first column corresponds to an ASE pedestal of $1.6 \kern0.1em {\times} \kern0.1em 10^{9} \kern0.2em \mathrm{W cm^{-2}}$ and the second column to $1.6 \kern0.1em {\times} \kern0.1em 10^{10} \kern0.2em \mathrm{W cm^{-2}}$, both for hydrogen. The third column corresponds to an ASE pedestal of $1.6 \kern0.1em {\times} \kern0.1em 10^{10} \kern0.2em \mathrm{W cm^{-2}}$ and carbon ions. The particle detection plane extends to $\pm 25 \kern0.2em \mathrm{mm}$, corresponding to a maximum divergence angle of $45 \degree$.}
  \label{fig:RCF22}
\end{figure}

\begin{figure}[!hb]
  \includegraphics[width=\linewidth]{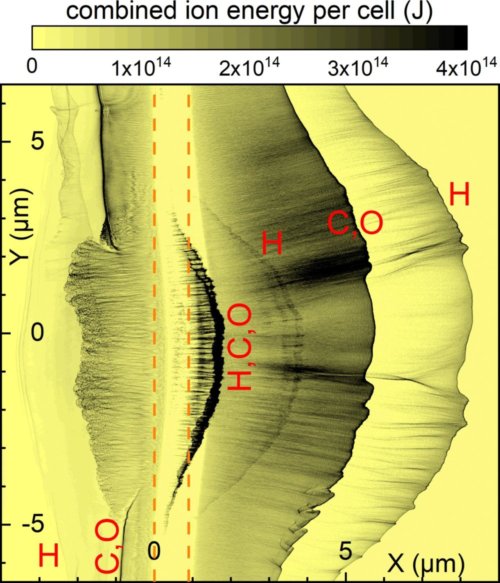}
  \caption{The combined energy of hydrogen, carbon and oxygen ions in each simulation cell for the 9th case of Fig. \ref{fig:pulse_particle}, at ${\sim} \kern0.1em 200 \kern0.2em \mathrm{fs}$ after the laser-foil interaction initiated. The red labels on the figure denote which particles are responsible for the localised condensations observed. The orange dashed lines represent the target surfaces prior the ASE pedestal irradiation.}
  \label{fig:energy_total}
\end{figure}

 The $\left( x , p_y \right)$ diagram forms a loop pattern which is a result of the proton beam divergence. As can be realised from Fig. \ref{fig:RCF20} and Fig. \ref{fig:RCF22} in combination with Fig. \ref{fig:energy_total}, protons originating from different positions of the target plane are emitted with a different divergence. The protons emitted from the centre of the focal spot region have almost no divergence, which increases as moving out of the focal spot; then, after the divergence has reached a maximum value, it starts decreasing asymptotically. Therefore, there exist two almost symmetric locations along $\bold{\hat{y}}$ where the divergence is maximised, which is directly connected with the $p_y$ magnitude, is seen by comparing the second column of  Fig. \ref{fig:RCF20} with the second and fourth rows of Fig. \ref{fig:PSh20}. As a result of that, the $\left( x , p_y \right)$ diagram forms a unique ring-pattern. This effect is observed for all foil thicknesses, although is more prominent for thinner foils where the divergence is further increased as a result of the initial target curvature, due to the act of the ASE pedestal. However, even the cases where the target surface is perfectly flat results on that divergence variation, meaning that it is a combination of both the foil surface field and the initial foil curvature.

\begin{figure}
  \includegraphics[width=\linewidth]{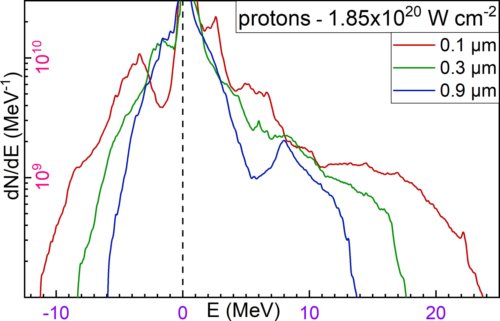}
  \caption{Proton spectra corresponding to $0.1 \kern0.2em \mathrm{\upmu m}$ (red line), $0.3 \kern0.2em \mathrm{\upmu m}$ (green line) and $0.9 \kern0.2em \mathrm{\upmu m}$ (blue line) thick Mylar targets for an ASE pedestal intensity of $1.6 \kern0.1em {\times} \kern0.1em 10^{10} \kern0.2em \mathrm{W cm^{-2}}$ and a main pulse intensity of $1.85 \kern0.1em {\times} \kern0.1em 10^{20} \kern0.2em \mathrm{W cm^{-2}}$. The ``negative'' energy axis denotes protons located at the target front region, while the positive energy axis denotes protons at the target rear region.}
  \label{fig:spectrum_hE20}
\end{figure}

\begin{figure}[!ht]
  \includegraphics[width=\linewidth]{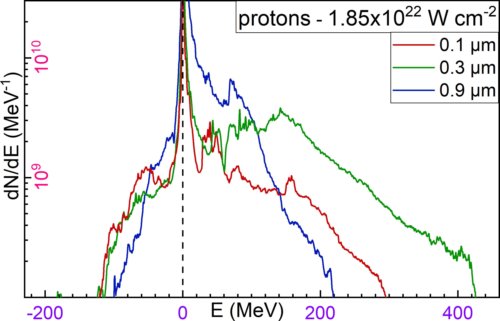}
  \caption{Proton spectra corresponding to $0.1 \kern0.2em \mathrm{\upmu m}$ (red line), $0.3 \kern0.2em \mathrm{\upmu m}$ (green line) and $0.9 \kern0.2em \mathrm{\upmu m}$ (blue line) thick Mylar targets for an ASE pedestal intensity of $1.6 \kern0.1em {\times} \kern0.1em 10^{10} \kern0.2em \mathrm{W cm^{-2}}$ and a main pulse intensity of $1.85 \kern0.1em {\times} \kern0.1em 10^{22} \kern0.2em \mathrm{W cm^{-2}}$. The ``negative'' energy axis denotes protons located at the target front region, while the positive energy axis denotes protons at the target rear region.}
  \label{fig:spectrum_hE22}
\end{figure}

\par Interestingly, Fig. \ref{fig:RCF20} and Fig. \ref{fig:RCF22} reveal an imperfectly symmetric pattern of the proton/ion distribution. The asymmetry is more prominent for the higher energy protons, where a deviation of the target normal axis is observed. This offset is higher for $0.1 \kern0.2em \mathrm{\upmu m}$ targets, where it has a value of ${\sim} \kern0.1em 15 \degree$ with a direction same as that of the laser pulse. This offset is believed to be caused by the altered target surface due to the ASE pedestal, where the main pulse faces a curved surface where a new target normal direction can be defined. This offset has also been demonstrated by Refs. \onlinecite{Yogo2008, Ogura2012}.


\par In order to obtain a quantitative understanding of the accelerated protons and ions behaviour, it is essential to extract spectra corresponding to both the front and rear foil surface. When protons are measured experimentally (for example with a radiochromic film \cite{Borghesi2002, Borghesi2003, Quinn2009}), the detector sees protons emitted only by the chosen foil surface with a certain divergence, as seen by the detection planes in Fig. \ref{fig:RCF20} and Fig. \ref{fig:RCF22}. In the figures, the distribution of both protons and carbon ions are shown; the oxygen ions are not presented because their distribution overlaps with that of carbon ions, since both ions have a charge to mass ratio of $0.5$.

\par In our simulations the proton distribution forms a gap on the lower energy region, which overlaps by the distribution of heavier ions; this behaviour can be visualised by comparing the second and third columns of Fig. \ref{fig:RCF20} and Fig. \ref{fig:RCF22}. The lowering of the proton spectra can be also seen in Fig. \ref{fig:spectrum_hE20} and Fig. \ref{fig:spectrum_hE22}, where a valley is formed in the spectra. This spectrum behaviour is explained by a volumetric effect of the 2D simulations, where the laser focal spot is represented by an infinitely long Gaussian along $\bold{\hat{z}}$, in contrast with a realistic 3D Gaussian spot, where the protons originating from the highest intensity regions are significantly overestimated.

\par The infinitely long intensity along $\bold{\hat{z}}$ also causes an overestimation of the total number of particles. However, a more realistic spectrum can be extracted (not in shape but rather in the total particles' number) if one assumes that most of the particles are emitted by a certain target area, which we assume to be approximately $5 \kern0.2em \mathrm{\upmu m}$. Furthermore, in both 2D and 3D simulations the particles' number density is the same, $N_{2D}=N_{3D}$. By definition, the number density in 2D is the number of particles ($n_{2D}$ for 2D and $n_{3D}$ for 3D) over the area, while in 3D is over the volume. If in both cases by defining a thickness parameter, $L$, and a spot radius, $R$, then the effective area is given by $2 \kern0.1em R \kern0.1em L$, while the effective volume by $\pi \kern0.1em R^2 \kern0.1em L$. By substituting the above definitions into $N_{2D}=N_{3D}$ gives:
\begin{equation}
 n_{3D} = \frac{\pi}{2} \kern0.1em R \kern0.1em n_{2D}
\label{rescale}
\end{equation}
As in our $5 \kern0.1em \mathrm{\upmu m}$ assumption, the 2D spectrum must be divided by a factor of ${\sim} \kern0.1em 1.27 \kern0.1em {\times} \kern0.1em 10^5$. It should be emphasised that this correction factor does not give an equivalent 3D spectrum but it rather rescales the 2D spectrum in a realistic number of particles. No detailed slope analysis is made for the spectra since only a 3D simulation would give a meaningful result; as a reference, for the lower intensity case the temperature is a few $\mathrm{MeV}$ and for the higher intensity it is in the range of tens of $\mathrm{MeV}$.

\begin{figure}[!ht]
  \includegraphics[width=\linewidth]{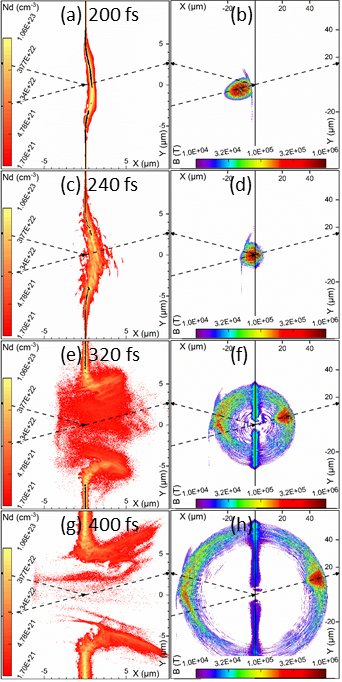}
  \caption{(left column) The time evolution of the electron number density. Densities higher than the relativistically corrected critical density are represented by the black area and those with lower than the classical critical density by white. (right column) Time evolution of the magnetic field. The figure corresponds to a $0.1 \kern0.2em \mathrm{\upmu m}$ thick foil with an ASE pedestal intensity of $1.6 \kern0.1em {\times} \kern0.1em 10^{9} \kern0.2em \mathrm{W cm^{-2}}$ and a main pulse intensity of $1.85 \kern0.1em {\times} \kern0.1em 10^{22} \kern0.2em \mathrm{W cm^{-2}}$.}
  \label{fig:evolution}
\end{figure}

\par As observed by  Fig. \ref{fig:spectrum_hE20}, the target front surface emits protons of significantly lower energy compared to the rear, in agreement with the TNSA mode. Furthermore, the maximum proton energy is not far from experimental observations under similar conditions \cite{Margarone2012}, where a detection threshold of $10^9$ corresponds to ${\sim} \kern0.1em 5 \kern0.2em \mathrm{MeV}$ for a $0.9 \kern0.2em \mathrm{\upmu m}$ thick Mylar target. Furthermore, from the spectra comparison one can observe that among the three thicknesses simulated, the $0.1 \kern0.2em \mathrm{\upmu m}$ results in a significantly higher proton energy. This observation is valid for both protons emitted from the target front and rear surfaces, as well as for ions (although with significantly lower energy per nucleon compared to hydrogen).

\par Similarly, Fig. \ref{fig:spectrum_hE22} shows a significantly lower proton energy for protons emitted from the foil front surface. However, in contrast with the lower intensity case, the front foil proton spectra does not vary significantly by altering the foil thickness. The protons originating from the target rear surface exhibit a more complex spectrum, which initially increases by reducing the foil thickness and after peaking at ${\sim} \kern0.1em 3 \kern0.2em \mathrm{\upmu m}$ it then drastically decreases. This inconsistency is explained by considering the level of the relativistically corrected critical density for the high intensity case (see Fig. \ref{fig:preplasma_lineout_15deg}). The reduction of the maximum proton energy is a result of the reduced target density for the case of the $0.1 \kern0.2em \mathrm{\upmu m}$ foil, which is ${\sim} \kern0.1em 4.83 \kern0.1em {\times} \kern0.1em 10^{22} \kern0.2em \mathrm{cm}^{-3}$, compared to a relativistically corrected critical density of $1.06 \kern0.1em {\times} \kern0.1em 10^{23} \kern0.2em \mathrm{cm}^{-3}$. As a result, the higher amplitude part of the laser pulse sees the target as transparent, resulting in a laser-foil interaction corresponding to an intensity of ${\sim} \kern0.1em 3.86 \kern0.1em {\times} \kern0.1em 10^{21} \kern0.2em \mathrm{W cm^{-2}}$, or a peak electric field with approximately half amplitude than the one initially assumed.

\par The preplasma effect on maximum proton energy becomes more obvious for lower ASE pedestal intensities, where the initial target density is less decreased. There, for the $0.1 \kern0.2em \mathrm{\upmu m}$ foil the electron number density is slightly above the relativistically corrected critical density (see Fig. \ref{fig:preplasma_lineout_15deg}). The regions where the target density is higher than the relativistically corrected critical density are shown with black colour on the left column of Fig. \ref{fig:evolution}. Although at a simulation time of $200 \kern0.2em \mathrm{fs}$ the target is not transparent, after the interaction with the main laser pulse the target starts expanding and its' density drops, resulting in a gradual transition to a transparent target, visualised by the evolution of the number density as shown in Fig. \ref{fig:evolution}.

\par In that case, the proton energy is enhanced by the Magnetic Vortex Acceleration \cite{Bulanov2007, Yogo2017} (MVA) mechanism. The basic requirement of the MVA mechanism is a target electron number density equal (or near) the critical number density, where the pulse is able to channel through the target, as shown on the right side of Fig. \ref{fig:evolution}; a small portion of the pulse is reflected, but $81 \kern0.2em \%$ of its' initial energy manages to pass through. As a result of the channelling pulse, a longitudinal electric field is created. Furthermore, electrons exhibit vortex trajectories which are associated with the generation of a quasi-static magnetic field. The magnetic field sustains the longitudinal electric field for longer times, benefiting the ion acceleration. Similar observations on optimised ion acceleration for the case of a near-critical density target can be found in Refs. \onlinecite{Matsukado2003, Yogo2008, Ogura2012, Esirkepov2014}.



\begin{figure}[!t]
  \includegraphics[width=1.0\linewidth]{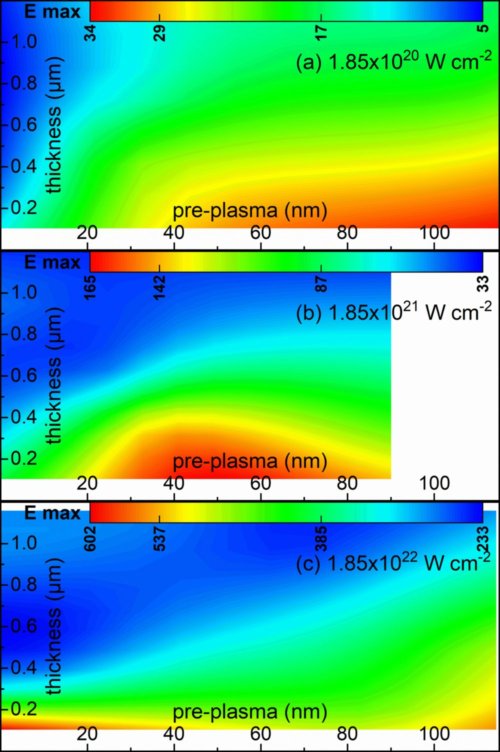}
  \caption{The maximum proton energy as a function of the preplasma scale-length and the effective foil thickness, for a main pulse intensity of a) $1.85 \kern0.1em {\times} \kern0.1em 10^{20} \kern0.2em \mathrm{W cm^{-2}}$, b) $1.85 \kern0.1em {\times} \kern0.1em 10^{21} \kern0.2em \mathrm{W cm^{-2}}$ and $1.85 \kern0.1em {\times} \kern0.1em 10^{22} \kern0.2em \mathrm{W cm^{-2}}$.}
  \label{fig:scalelength_contour}
\end{figure}

\par The effect of the preplasma formation is visually summarised in Fig. \ref{fig:scalelength_contour}, where the maximum proton energy is plotted as a function of the preplasma scale-length and the effective foil thickness. The preplasma scale-length is measured as the difference of the spatial locations corresponding to the relativistically corrected critical density and the relativistically corrected critical density of half the maximum electric field value. The effective thickness is measured as the foil thickness corresponding to the relativistically corrected critical density. Furthermore, both the preplasma scale-length and foil thickness have been measured with respect to the $15 \degree$ laser incidence angle.

\par Similar contours to those presented in Fig. \ref{fig:scalelength_contour} are presented in Ref. \onlinecite{Esirkepov2014}. The main difference of the two cases is the way the preplasma scale-length is measured, where in our representation only the higher half of the electric field is considered on the estimation of the scale-length. In that region the preplasma distribution is approximately linear, meaning that the scale-length can be also seen as a well-defined preplasma gradient. Another significant difference is that in our case no flat-top profile exists in the electron number density distribution (therefore no plateau region exists) due to the significantly thin targets examined, strongly altered by the ASE pedestal. Fig. \ref{fig:scalelength_contour}(a) (intensity of $1.85 \kern0.1em {\times} \kern0.1em 10^{20} \kern0.2em \mathrm{W cm^{-2}}$) reveals that at a preplasma scale-length of ${\sim} \kern0.1em 110 \kern0.2em \mathrm{nm}$ the maximum proton energy is significantly enhanced for all the foil thicknesses examined. However, the maximum proton energy is given by thinner ($0.1 \kern0.2em \mathrm{\upmu m}$) targets.

\par However, the effect of proton energy increase by increasing the preplasma scale-length becomes less significant as the intensity is increased, as can be observed by comparing Fig. \ref{fig:scalelength_contour}(a) with Fig. \ref{fig:scalelength_contour}(b) and Fig. \ref{fig:scalelength_contour}(c). As shown in Fig. Fig. \ref{fig:scalelength_contour}(b), an optimum preplasma scale-length of ${\sim} \kern0.1em 45 \kern0.2em \mathrm{nm}$ exists, which as in the case of lower intensity, the proton energy is maximised for thinner targets. Finally, Fig. \ref{fig:scalelength_contour}(c) reveals a completely different behaviour since the preplasma existence little affects the maximum proton energy obtained. Again as in the case of lower intensity, thinner targets result in a significantly higher proton energy. For this intensity regime, the  thinner targets benefit the proton acceleration as explained in Fig. \ref{fig:evolution}, where although the targets are initially opaque, their thin electron density distribution falls below critical soon after the interaction with the main pulse initiates. Therefore, at high intensities and very thin targets the existence of a preplasma can be problematic requiring lasers with extremely high contrast for an optimum proton acceleration.


\section{Summary and Conclusions} \label{Conclusions}

\par The main aim of this work is to investigate the effect of a preplasma on proton acceleration. An initially steep density gradient Mylar foil is assumed, which then interacts with a $1 \kern0.2em \mathrm{ns}$ long ASE pedestal. The effect of the ASE pedestal on the modification of the initial density distribution is studied by a hydrodynamic code, which revealed a significantly altered density distribution. The new distribution is characterised by an exponential-like decreasing preplasma distribution, which becomes approximately linear in the most dense parts of the distribution. Furthermore, in the focal spot region the initially flat-foil is significantly curved, where the curvature has been more prominent for thinner foils.

\par Once the main laser pulse reaches the most dense regions of the electron number density distribution it then transfers part of its energy into hot electrons. The lower part of the pulse interacts as soon as the classical critical density is reached and the higher part interacts at the relativistically corrected critical density. A small fraction of the pulse is trapped by the target, driving an electron population along the target surface. The rest of the pulse is either reflected or transmitted through the target, depending on whether or not the target survives the interaction with the main pulse. This fraction of the initial pulse remains highly localised and drives a portion of the electron population into significantly higher energies.

\begin{figure}
  \includegraphics[width=\linewidth]{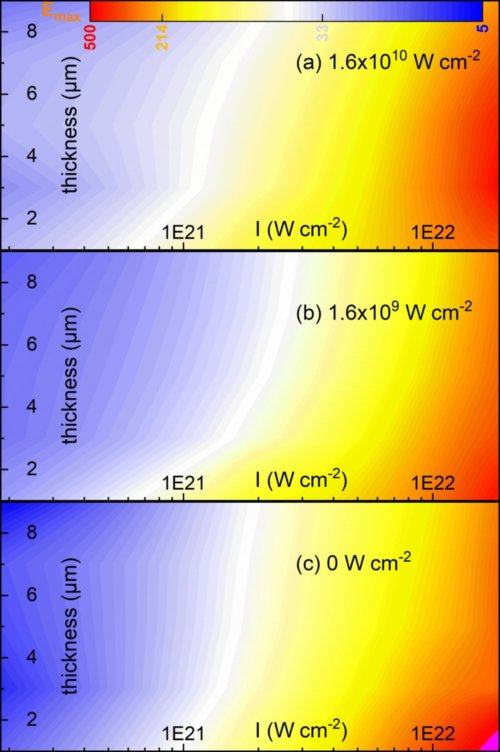}
  \caption{A collective comparison of the maximum proton energy (in logarithmic scale) for the three foil thicknesses presently examined and the logarithm of the main pulse intensities, for an ASE pedestal intensity of a) $1.6 \kern0.1em {\times} \kern0.1em 10^{10} \kern0.2em \mathrm{W cm^{-2}}$, b) $1.6 \kern0.1em {\times} \kern0.1em 10^{9} \kern0.2em \mathrm{W cm^{-2}}$, c) $0 \kern0.2em \mathrm{W cm^{-2}}$.}
  \label{fig:maxE}
\end{figure}

\par Three foil thicknesses are examined ($0.1 \kern0.2em \mathrm{\upmu m}$, $0.3 \kern0.2em \mathrm{\upmu m}$ and $0.9 \kern0.2em \mathrm{\upmu m}$), in combination with intensities of $1.85 \kern0.1em {\times} \kern0.1em 10^{20} \kern0.2em \mathrm{W cm^{-2}}$ and $1.85 \kern0.1em {\times} \kern0.1em 10^{22} \kern0.2em \mathrm{W cm^{-2}}$, and further combined with three ASE pedestal intensities of $0 \kern0.2em \mathrm{W cm^{-2}}$, $1.6 \kern0.1em {\times} \kern0.1em 10^{9} \kern0.2em \mathrm{W cm^{-2}}$ and $1.6 \kern0.1em {\times} \kern0.1em 10^{10} \kern0.2em \mathrm{W cm^{-2}}$. In all cases the $0.1 \kern0.2em \mathrm{\upmu m}$ thick targets give the highest proton energy apart a single case of ASE pedestal intensity of $1.6 \kern0.1em {\times} \kern0.1em 10^{10} \kern0.2em \mathrm{W cm^{-2}}$ and main pulse intensity of $1.85 \kern0.1em {\times} \kern0.1em 10^{22} \kern0.2em \mathrm{W cm^{-2}}$. The reason of the lower proton energy for that case is that the target becomes transparent for the higher amplitude of the main pulse due to the significantly lowered peak electron number density after the ASE pedestal interaction.

\par For a main pulse intensity of $1.85 \kern0.1em {\times} \kern0.1em 10^{20} \kern0.2em \mathrm{W cm^{-2}}$ the effective preplasma scale-length (as defined in the text) is found to have an optimum value at ${\sim} \kern0.1em 110 \kern0.2em \mathrm{nm}$ (which is the maximum value examined in this work), where the energy increase (compared to the case of no preplasma) is approximately $7$ times; longer scale-lengths can potentially give even longer proton energies. However, at $1.85 \kern0.1em {\times} \kern0.1em 10^{22} \kern0.2em \mathrm{W cm^{-2}}$ an energy increase of approximately $2$ times is achieved with a scale-length of ${\sim} \kern0.1em 110 \kern0.2em \mathrm{nm}$. On the other hand, at that intensity level the $0.1 \kern0.2em \mathrm{\upmu m}$ thick targets give slightly higher energies when no preplasma is present. The outcome of all cases simulated are compared in Fig. \ref{fig:maxE}, where the maximum proton energy is potted as a function of the initial foil thickness and the peak intensity. the data are grouped in three categories regarding the ASE pedestal intensity. A general observation is that although targets without preplasma give significantly lower proton energies for lower intensities, this difference cease at higher intensities.

\begin{acknowledgments}
The authors thank T.M. Jeong and D. Margarone for fruitful discussions. This work is supported by the project High Field Initiative (CZ.02.1.01/0.0/0.0/15\_003/0000449) from the European Regional Development Fund. It was also in part funded by the UK EPSRC grants EP/G054950/1, EP/G056803/1, EP/G055165/1 and EP/ M022463/1.
\end{acknowledgments}

\appendix

\nocite{*}
\bibliography{bibliography}

\end{document}